\documentclass[epj]{svjour}
\usepackage{graphicx}
\usepackage{amsmath}
\usepackage{amssymb}
\usepackage{braket}
\usepackage{bigints}
\usepackage[table]{xcolor}
\usepackage{cuted}

\newenvironment{widetext}{%
  \begin{strip}
  \vskip-1.5ex
  \noindent\rule{0.45\textwidth}{0.4pt}
  \vskip1ex
}{%
  \vskip1ex
  \noindent\hspace{0.55\textwidth}\rule{0.45\textwidth}{0.4pt}
  \vskip-1.5ex
  \end{strip}
}

\begin{document}
\title{Glueball interactions from the colour Van der Waals potential}
\author{Juan C\'anovas C\'uneo   \and Felipe J. Llanes-Estrada 
}                     
\institute{Departamento de F\'{i}sica Te\'orica \& IPARCOS, Facultad de CC. F\'{i}sicas, Plaza de las Ciencias 1, 28040 Madrid, Spain}
\date{\today}
%
\abstract{
We examine pure Yang-Mills theory glueball-glueball interactions in constituent-gluon approaches. Because of the large mass gap, Van der Waals interactions are relatively more significant than in light-quark QCD hadrons where the pion gives rise to strong Yukawa exchanges.
We find that the colour Van der Waals potential, computed along the traditional lines of the quantum London-Eisenschitz-Wang force, is a relevant interaction at distances between about 0.66 fm (when the glueballs are in contact) to 1.9 fm (when the virtual string tension among the colour-polarized glueballs in the intermediate state breaks down). In employing a Cornell potential as the microscopic one among colour charges, we find that the Van der Waals interaction derived from the linear potential part closely cancels that of the Coulomb one for the ground state glueballs. This means that the interaction, while sizeable and capable of saturating the lattice data depending on parameters, is far weaker than the London force derived from the Coulombic part alone.
(This suggests that one should move effective theories for glueballs beyond dilaton-type approaches and perhaps deploy some variant of Van der Waals Effective Field Theory.)   
\PACS{
      {11.15.-q}{Gauge field theories} 
     } 
} 
\maketitle
\section{Introduction}  \label{intro}
Nucleon-nucleon interactions~\cite{Epelbaum:2008ga} are dominated, at long range, by one-pion and at intermediate range,
correlated two-pion ($\sigma$) exchanges. At yet shorter ranges, nuclear forces are sometimes
parametrized by heavier-meson and quark-exchange interactions, substituted for contact terms in the
spirit of effective field theories which are agnostic about microscopic mechanisms.

In the quarkless ``pure'' Yang-Mills theory~\footnote{The Lagrangian is the well-known
$ \mathcal{L}_{YM}=\frac{-1}{4}F^a_{\mu\nu}F^{a\mu\nu}\ . $}, the bound and resonant states are glueballs or oddballs~\cite{Wiedner:2026mqv,Llanes-Estrada:2021evz} and  
pion exchange is unavailable, so the singlet particle (lightest scalar glueball) exchanges are also of short range
(as $M_{gg}\sim 1.7$ GeV instead of $M_\pi\sim 0.138$ GeV). 
Thus, exchange interactions compete with other microscopic mechanisms. While at the shortest ranges the forces manifest in the QCD Lagrangian presumably dominate, in the intermediate range Van der Waals interactions, which are residual from colour potentials~\cite{Gavela:1979zu}, might make a contribution. 

\begin{figure}
    \centering
        \includegraphics[width=1.1\linewidth]{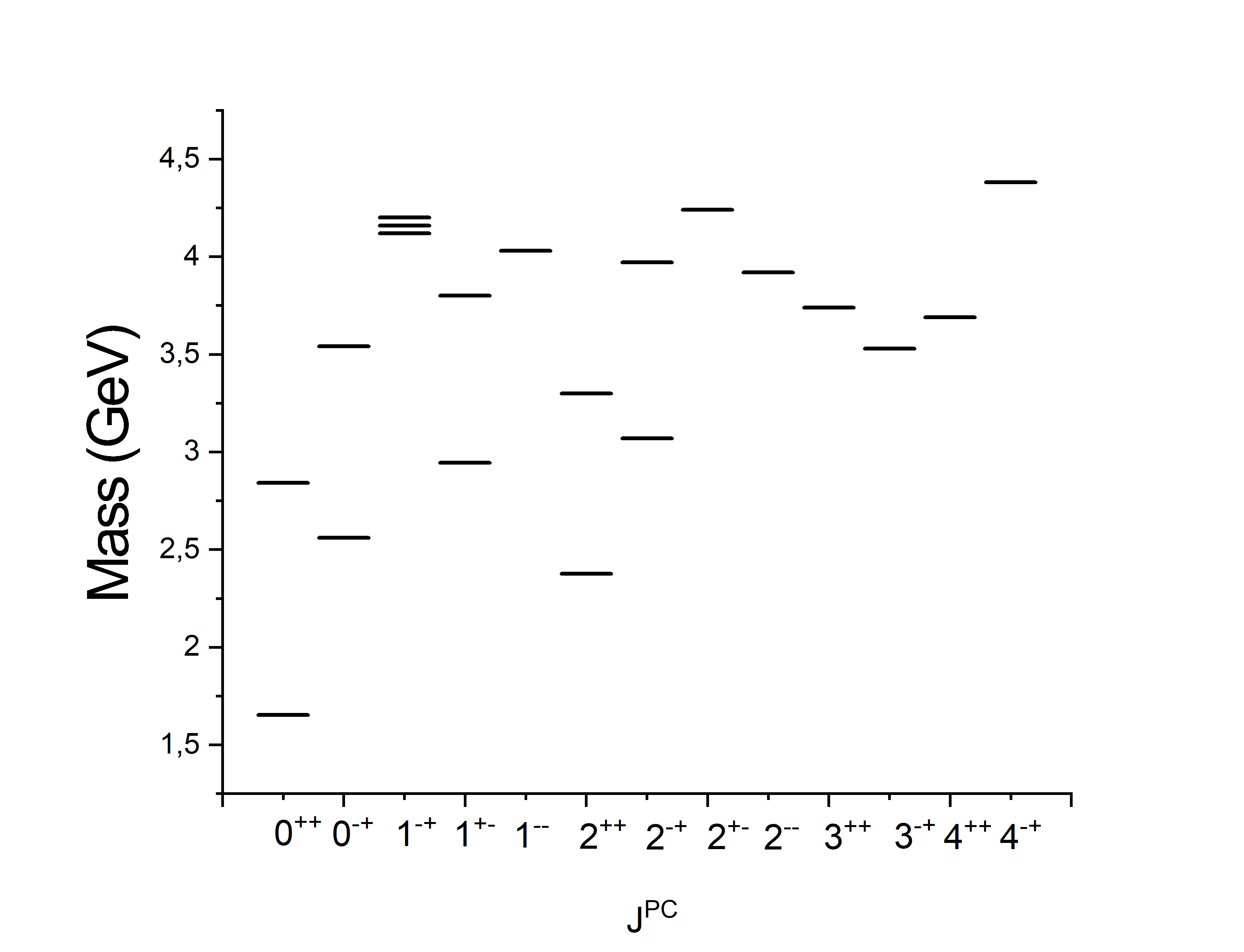}
        \caption{Lattice QCD data~\cite{Athenodorou:2020ani} of the pure Yang-Mills spectrum (basically, the glueball and oddball spectrum) The YM mass-gap (mass of the lowest eigenstate, the scalar glueball on the lower left corner) must survive glueball-glueball interactions.} 
        \label{glueball_spectrum}
\end{figure}

We see two reasons to revisit Van der Waals interactions, here in the context of pure Yang-Mills theory (for which they have not yet been treated). One is to understand glueball-glueball interactions to ensure that they are not intense enough to affect the Yang-Mills mass gap -see Fig.~\ref{glueball_spectrum}- (whose understanding is part of a relevant Clay Mathematics Institute Millennium Prize Problem). The other is the proliferation of dark sectors which include a Yang-Mills component~\cite{Huang:2020crf,Yamanaka:2019aeq,Morgante:2022zvc}.

Because in confining theories gluon exchange is energetically forbidden at long distances, Van der Waals and colour-singlet Yukawa exchanges are the interactions of interest.

Thus, we have undertaken an investigation into the residual Van der Waals interaction due to colour forces involving only gluons. 

The overall outcome of the analysis seems clear: Van der Waals interactions in pure Yang-Mills theory are  significant, unlike among nucleons in full Quantum Chromodynamics.

\subsection{Interplay of the Yang-Mills mass gap, the Yukawa and the Van der Waals interaction}
 \label{subsec:incompatibilidadVdW}
A way to think of the potential in an interacting quantum field theory is 
 \begin{align}
     V(r)=-\frac{g^2}{4\pi r}\int_0^\infty\frac{d\mu^2}{2\pi}\rho(\mu^2)e^{-\mu r}
     \label{Yuk}
 \end{align}
in terms of the spectral density  $\rho(\mu^2)$  which appears in the full Feynman propagator's Källén-Lehmann \cite{Peskin:1995ev} representation
\begin{equation}
    \hat{D}_T(p)=\int_0^\infty\frac{d\mu^2}{2\pi}\frac{i\rho(\mu^2)}{p^2-\mu^2+i\epsilon}\ ;
\end{equation} 
this is because in the nonrelativistic approximation, the potential representing the exchange of this particle is
\begin{equation}
V(r)=-g^2\int\frac{d^3\boldsymbol{p}}{(2\pi)^3}e^{i\boldsymbol{p}\cdot\boldsymbol{r}}\hat{D}_T(p^0=0,\boldsymbol{p})\ .
\end{equation}

In a gapped theory such as we believe Yang-Mills to be,  $\exists M>0| \rho(\mu^2)=0\ \forall \mu\in[0,M)$; 
this means that the spectral integration runs over $[M,\infty)$. 

In practice, because for large $r\rightarrow\infty$ the exponential factor $e^{-\mu r}$ of Eq.~(\ref{Yuk}) is of rapid decrease with increasing $\mu$, the upper integration limit may be truncated to a certain $\mu_0$, closer to $M$ for larger $r$.

Then, asymptotically, the potential of Eq.~(\ref{Yuk}) takes the short-range Yukawa form $V\propto e^{-M r}/r$ discussed later in subsec.~\ref{subsec:dilaton} (see for example Eq.~(\ref{Yukawapotential})).
This is unlike the Van der Waals potential of Eq.~(\ref{Poteff}), a (negative) power law. For example, for the Cornell potential we will remind the reader later on that ($V\sim\frac{1}{r^3}$) due to the long-distance linear growth of the underlying potential.

The situation is quite ironic. If the theory has no mass gap (as QCD in the chiral limit), the spectral argument permits asymptotic Van der Waals interactions, but massless-Goldstone boson (pion) exchange then yields a Coulomb potential which is much longer range, and the dominant interaction. 

If the theory on the other hand presents a mass gap, as is the case of the pure Yang-Mills theory here discussed, then Van der Waals interactions cannot be the asymptotic ones. Nevertheless, they can still dominate overall due to larger strength in the short- and middle- distance range (because of the $e^{-Mr}$ suppression in the Yukawa exchange with a large mass gap), and this is what we here investigate.

\subsection{Scales relevant in $SU(3)$ Yang-Mills theory \& QCD}
\label{subsec:scales}

The interaction between coloured gluon clusters proceeds by gluon exchange yielding colour forces. 
If we insist that the gluon clusters be colour-singlet glueballs, then singlet  (induced dipole-induced dipole) Van der Waals interaction can take place.  This happens to a distance at which the value of the relevant potential in the dipole-excited channel equals the mass gap. At that point, breaking the gluon string is favorable and the potential transitions to a Yukawa one.  

At what distance should the transition from Van der Waals to Yukawa interaction take place?
Let us equate the energy in the chromoelectric flux tube from the Cornell potential (virtually appearing in the Van der Waals interaction) to the mass gap,
\begin{equation}\label{HowFarVdW}
a |{\bf R}^{\rm YM}|_{\rm VdW}-\frac{\alpha_s}{|{\bf R}^{\rm YM}|_{\rm VdW}}=m_{\rm gap}\ .
\end{equation}
Typical values~\cite{Llanes-Estrada:2005bii,Concejo:2023xzf}~$\alpha_s=0.4$, $a=0.18\  \rm{GeV}^2$ and $m_{\rm gap}=1.7\ \rm{GeV}$ 
yield a characteristic distance
\begin{equation}\label{YMVdWrange}
|{\bf R}^{\rm YM}|_{\rm VdW}\simeq 1.9\ \rm{fm}\ .
\end{equation}

It is interesting to compare this distance with the Yukawa one, the inverse of the mass gap,
\begin{equation} \label{GlueballYukawa}
|{\bf R}^{YM}|_{\rm Yukawa}=\frac{1}{m_{\rm gap}}\simeq0.12\ \rm{fm}
\end{equation}
This is much smaller! Although, as argued in subsection~\ref{subsec:incompatibilidadVdW}, the asymptotic potential is the Yukawa glueball exchange, by the time the Van der Waals-like interaction shuts off, that Yukawa potential is already several characteristic lengths out, and is considerably smaller, with a suppressing factor
\begin{equation}
  \exp(-m_{\rm gap}|{\bf R}^{\rm YM}|_{\rm VdW})\simeq 7.2 \cdot 10^{-8}  
\end{equation}
so there is a chance that the Van der Waals type interaction is competitive. 

It remains to establish the small-$|{\bf R}|$ validity of the Van der Waals potential. At very short distances, less than twice the average glueball radius $\langle\rho\rangle$, the gluon clusters overlap forming a 4-gluon glueball. Therefore, the interval where it makes sense to speak of the VdW potential is $[2\langle\rho\rangle,|{\bf R}^{\rm YM}|_{\rm VdW}]$
\medskip

Let us then turn to a quick estimate of the energies involved by looking at the order of magnitude of the interaction potential.

Although the detailed calculation will follow later on, it will be useful to think now in terms of Eq.~(\ref{Poteff}) below.
The order of magnitude of the transition between states in the numerator of the Van der Waals potential can be approximately taken as 
\begin{equation}
\left(\frac{a}{R}-\frac{\alpha_s}{R^3}\right)^2||\boldsymbol{D}||^4\ .
\end{equation} 
The order of magnitude of the dipole is the colour-charge unit times the average radius 
\begin{equation}
||\boldsymbol{D}||~ g_s\langle\rho\rangle=4\pi\alpha_s\langle\rho\rangle\ .
\end{equation}
On the other hand, the suppressing denominator between the singlet-singlet (subindex 1) and the virtual coloured-coloured states (subindex $\chi$) is of order
$2(E_1-E_\chi)-(aR-\frac{\alpha_s}{R})$
To excite the glueball from a singlet to another representation (obviously, not in isolation, which would be divergent, but as a part of a larger, overall singlet system) would cost an energy of the order of half the mass gap (one ``constituent gluon'' needs to be emitted and its mass shared between both clusters), $(\simeq m_{\rm gap}/2)$.

Thus, very sketchily, we can take $2(E_1-E_\chi)\simeq m_{\rm gap}$. 
Putting both together, the order of magnitude of the potential is then
\begin{equation}
|V_{eff}|\sim \frac{(aR-\frac{\alpha_s}{R})^2\ (4\pi\alpha_s)^4\ (\langle\rho\rangle/R)^4}{m_{\rm gap}-(aR-\frac{\alpha_s}{R})}\ .
\end{equation}

We can now evaluate it at the lower limit of the validity range $R=2\langle\rho\rangle$, 
with the numerical figures for $a$, $\alpha_s$ and $m_{\rm gap}$ already adopted, and with  $\langle\rho\rangle=0.33\rm{fm}$.
This value~\cite{Shuryak:2026grt} for the size of the $0^{++}$ scalar glueball is quite small as hadrons go, making it a very compact object (our own calculation, within 5\% of this, will be shown below in Eq.~[\ref{radiuscalc}]). 
The resulting potential energy at glueball-glueball contact is then estimated as 
\begin{equation}
    |V_{eff}(2\langle\rho\rangle)|\sim 80\ {\rm MeV}\ . 
\end{equation}
An immediate consequence is that this potential is too feeble to erase the Yang-Mills mass gap (at 1.7 GeV), which is in accordance with lattice measurements.

\subsection{Effect of quarks}

Although this article is concerned with pure Yang-Mills theory,
quickly comparing with the Van der Waals interaction in full QCD, where it is known to be of little relevance, is in order.

Let us rethink the window where this London force is of interest, but now in the presence of light quarks. 
In QCD with fermions, upon reducing the mass gap to the pion mass due to its nature of a quasiGoldstone boson of spontaneous chiral symmetry breaking, the colour string in the intermediate virtual state extends a far smaller distance. Again equating the Cornell 
potential as in Eq.~(\ref{HowFarVdW}), but now to the pion mass $m_\pi=140\  \rm{MeV}$, an interaction range for the Van der Waals interaction is obtained,
\begin{equation} \label{VdWQCD}
|{\bf R}^{\rm QCD}|_{\rm VdW}=0.38\ {\rm fm} \ .
\end{equation}
Compare this with Eq.~(\ref{YMVdWrange}): it is five times smaller!
Such range is even below the pion charge radius 
at $0.6\ \rm{fm}$~\cite{ParticleDataGroup:2024cfk} (and those of many other hadrons, such as the nucleon). There is no comparable clean-distance window in QCD with light quarks where the colour Van der Waals force can make a contribution, as has been long known~\cite{Greensite2011}.

\begin{figure}
    \centering
    \includegraphics[width=\linewidth]{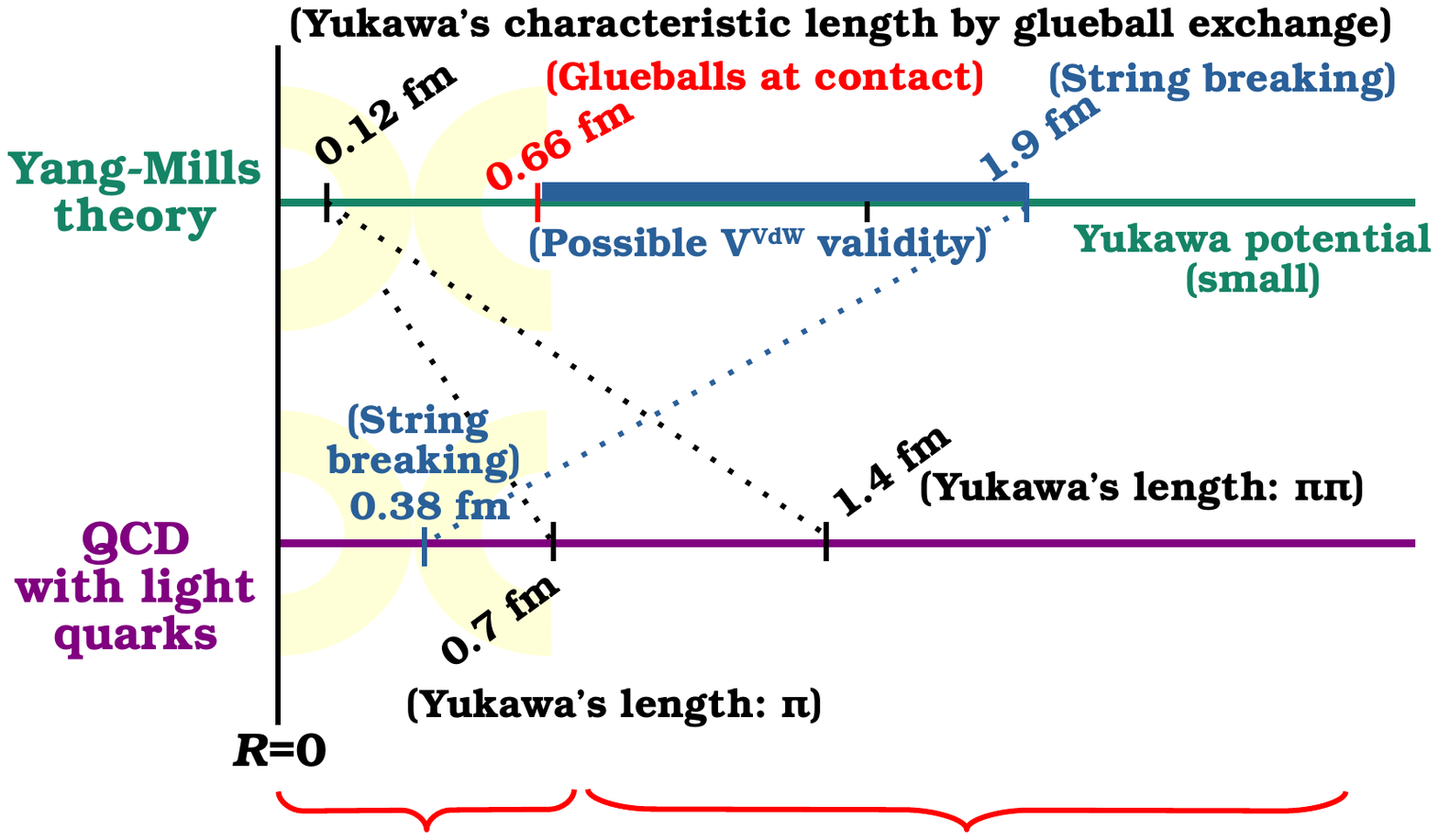}
    \caption{Various scales useful to understand why we think that colour Van der Waals interactions may play an important role in the glueball-glueball potential while they are quite irrelevant in QCD, with light fermions added. Indeed, in the first case (upper line) the virtual string-breaking length is larger than twice the glueball radius, opening a window (thick blue line between 0.66fm and 1.9fm) where the London force may be sizeable because Yukawa's glueball-exchange is exponentially suppressed.  The lower line shows that, in the presence of light pions, the scales are inverted and there is no Van der Waals window.}
    \label{fig:scales}
\end{figure}

By the way, the Yukawa interaction due to pion exchange spans the characteristic range of the strong nuclear force
\begin{equation}
   |{\bf R}^{\rm QCD}|_{\rm Yukawa} \sim \frac{1}{m_\pi} \simeq 1.4\ {\rm fm}\ .
\end{equation}
Thus, in no way our discussion of subsec.~\ref{subsec:scales} contradicts traditional understanding of nucleon-nucleon interactions. There simply is no range at which the Yukawa potential is small and there is a chance for virtual strings to form in intermediate steps supporting a Van der Waals potential. The situation is depicted in Fig.~(\ref{fig:scales}).

We now turn to heavy quarks.
Double quarkonium is a system where colour-dipole interactions have indeed been reported~\cite{Brambilla:2010cs}  
and quarkonium-nucleon one where they may play a role at short distances, but they quickly give way to two-pion exchange~\cite{Lyu:2024ttm}
(which is Yukawa-like but with an additional power of $1/r$).

The average radius of the $J/\psi$ is $0.25\ \rm{fm}$~\cite{Brambilla:2010cs}, and its mass $3.096\ \rm{GeV}$\cite{ParticleDataGroup:2024cfk}, 
Two-pion exchange, with a minimum energy of 0.28 GeV, gives a characteristic Yukawa radius of 0.7 fermi (in a theory with only charm and the Yang-Mills fields we would revert to Eq.~(\ref{GlueballYukawa}) with a 0.12 fm range) and the string-breaking distance with two pions would yield double the number in Eq.~(\ref{VdWQCD}), or about 0.76 fm. Hence, our rule of thumb suggests that in the presence of heavy+light quarks, the Van der Waals interactions associated purely to gluodynamics is not dominant, but competes with Yukawa exchange at intermediate distances.
These potentials are known to be sufficient to form bound states/hadron molecules with energies of order $10$ MeV but not much more, and certainly not sufficient to alter the mass gap in any of the versions of the theory, with or without light quarks.

\newpage

\section{Few-body glueball representation} \label{sec:salto de masa}

To present our results for the glueball-glueball interaction in the Van der Waals range, we first need to 
briefly discuss their few-body representation, motivating from the fundamental underlying field theory the 
approximate Tamm-Dancoff eigenvalue problem which can be used to reduce them to a quantum mechanical formulation. 

First, how can one, from a theory of massless quanta, reduce the system's description to few (two, three...) particles?
This is thanks to the Yang-Mills mass gap itself. 

Adding a mass term to Maxwell's equations in a massless theory, for example via the Proca equation,
\begin{align}
         \partial_{\mu} F^{\mu\nu}+m^2A^{\nu}&=0
 \label{EcsProca}
\end{align}
leads to loss of gauge invariance. (To see it, combine Eq.~(\ref{EcsProca}) with the definition of
$F^{\mu\nu}$), obtaining $m^2\partial_{\mu}A^{\mu}=0$, and if the boson is massive $m\neq0$, 
the Lorentz gauge condition immediately appears  $\partial_{\mu}A^{\mu}=0$, so that gauge invariance is absent). 
This means that the gauge-invariant mass-gap in Yang-Mills theory must be assessed through colour-singlet, physical states, basically the glueball mass. Yet, if one is willing to work in a fixed-gauge representation such as Landau gauge~\cite{Huber:2025kwy} or Coulomb gauge~\cite{Szczepaniak:1995cw}, it is fine to use the concept of a mass-gap for the coloured gluon, in the understanding that it is an auxiliary quantity which cannot have physical meaning by itself, but it can be useful to obtain qualitative information about what is happening in the theory.

Additionally, dilatation and full conformal symmetry are lost due to the gluon mass gap~\cite{Cornwall:1981zr,Aguilar:2019kxz}, but the trace anomaly is already signaling that loss of symmetry.
In terms of $\beta_g$ (evolution of the coupling constant) and $\gamma_m$ (anomalous mass dimension of the fermions), this reads~\cite{Hoferichter:2025ubp}
\begin{align}
    \theta^{\mu}_{\mu\text{, YM}}=\frac{\beta_g}{2g}F_{\mu\nu}^aF_a^{\mu\nu}+\sum_{q=1}^{N_f}m_q(1-\gamma_m)\bar{\psi_q}\psi_q
\end{align}

Once a mass gap is formed, however, at given energy there is little or no phase space for higher Fock states, which become small virtual corrections: the low-lying glueballs are dominated by quasiparticles, which explains the success of such simple-minded potential models in interpreting Lattice-Gauge-Theory results.

\subsection{Mass-gap generation}
A practical dynamical constituent description which incorporates many known features of spontaneous mass generation is the 
North Carolina State  (NCSU) Coulomb Hamiltonian approach~\cite{Concejo:2023xzf,Szczepaniak:1995cw,Llanes-Estrada:2000ozq}. A simplified Hamiltonian for the gluon vector fields 
in Coulomb gauge is deployed which respects all (global) symmetries, particularly colour, although it is an approximate description and gauge invariance is sacrificed.

A gluon condensate populates the ground state, $|\Omega\rangle$. This is a vacuum for the quasigluon modes $\alpha$, $\alpha^\dagger$ obtained from the free-wave modes $a$, $a^\dagger$ via a hyperbolic Bogoliubov rotation, yielding the usual coherent state
\begin{equation}
| \Omega\rangle=\mathcal{N}e^{(-\int\frac{d^3\boldsymbol{k}}{2(2\pi)^3}\text{tanh}\theta^a_{\boldsymbol{k}}(\delta_{ij}-\hat{k}_i\hat{k}_j)a_i^{a^{\dagger}(\boldsymbol{k})}a_j^{a^{\dagger}(-\boldsymbol{k})})}|0\rangle\ .
\end{equation}

Applying the variational Rayleigh-Ritz principle to approximately minimize $\langle H\rangle_{\Omega}$ 
by varying the one-gluon dispersion relation
$E_{\boldsymbol{q}}^d$ yields a zero-momentum nonvanishing energy, which can then be interpreted as a Coulomb-gauge gluon mass (which is not gauge invariant, but is consistent within a fixed-gauge approach; an alternative parametrization is a Gribov-type formula with a divergent behaviour~\cite{Yepez-Martinez:2012gdo}, but the difference will be small in phenomenological applications). 

 Then, the resulting mass-gap equation just requires specifying the effective potential in the Hamiltonian,
 \begin{equation}
\hat{V}_{eff}(\boldsymbol{k},\boldsymbol{q})=\frac{1}{2\pi}\int d\Omega(\hat{V}(|\boldsymbol{k}-\boldsymbol{q}|)(1+(\hat{\boldsymbol{k}}\cdot\hat{\boldsymbol{q}})^2))
\end{equation}
where $\hat{V}(|\boldsymbol{k}-\boldsymbol{q}|)$ is the Fourier transform to momentum space of the (assumed central) colour-potential  $\hat{\boldsymbol{k}}$ and $\hat{\boldsymbol{q}}$ unit vectors parallel to   
$\boldsymbol{k}$ and $\boldsymbol{q}$.
\begin{align}
   \underbrace{( E_{\boldsymbol{q}})^2=\boldsymbol{q}^2}_{\text{Free\ Gluon}}-\frac{C_A}{4}\int_0^{\infty}\frac{dk}{(2\pi)^2}k^2\hat{V}_{eff}(\boldsymbol{k},\boldsymbol{q})\left( \frac{E_k^2-E_q^2}{E_k} \right)
   \label{gapmasa}
\end{align}

This is a nonlinear integral equation $E_{\boldsymbol{q}}$ 
which was numerically solved by iteration.
 A common choice for $V$  is the Cornell (Coulomb+linear) potential
\begin{equation}
V_{\rm int}(|\boldsymbol{x}-\boldsymbol{y}|)=-\frac{\alpha_s}{|\boldsymbol{x}-\boldsymbol{y}|}+ a|\boldsymbol{x}-\boldsymbol{y}|\ ,
\label{Cornellpot}
\end{equation} where $\alpha_s$ is the strong-interaction constant and the string tension is $ a=0.18\ \rm{GeV}^2$.

  Subtracting the equation in the MOM scheme to tame the UV  and regulating the IR behaviour of the linear potential which retains a logarithmic divergence yields the gapped gluon dispersion relation $E(\boldsymbol{q})$. That program will not be repeated here as it has been extensively reported elsewhere for QCD and for various other YM theories~\cite{Concejo:2023xzf}.
 The emerging gluon mass $m=E(0)$ grows with the dimension $N$ of the gauge group due to the Casimir $C_A=N$  factor in Eq.~(\ref{gapmasa}). 
 
 This approximation has the inconvenient that the potential is input to the integral equation on phenomenological grounds. Lifting this simplification entails writing integral equations for higher-point Green's functions, which has also been explored before~\cite{Szczepaniak:2001rg,Campagnari:2019zso}.

\subsection{Two- and three-gluon bound states}
The linear potential associated to the Coulomb NCSU Hamiltonian removes coloured states from the spectrum, lifting them to infinite mass, so that the remaining glueballs with finite energy must be colour singlets.

The Tamm-Dancoff Approximation (TDA) is applied to the two-gluon state in its centre of mass frame,  so that the variational state is 
\begin{equation}|J^{PC}\rangle=\int\frac{d\boldsymbol{p}}{(2\pi)^3}\chi^{JPC}_{ij}(\boldsymbol{p})a_i^{\dagger b}(\boldsymbol{p})a_j^{\dagger b}(-\boldsymbol{p})|\Omega\rangle 
\end{equation}.
The wavefunctions $\chi^{JPC}_{ij}$ satisfy a Schr\"odinger-like integral eigenvalue problem~\cite{Szczepaniak:1995cw}, again numerically solved. 
The resulting states were seen to be in very reasonable agreement with lattice computations for the positive charge-conjugation states, with an adequate 1.7 GeV mass gap to the vacuum, and the lightest glueball the scalar state $J^{PC}=0^{++}$
(the lattice estimate standing at 1.653(26) GeV~\cite{Athenodorou:2020ani}).

Negative $C$-parity states exceed the two-gluon wavefunction though, and require addressing a three-body problem (just this qualitative observation naturally explains why they are more massive in the lattice evaluation); a more quantitative variational estimate was already carried out in the past~\cite{Llanes-Estrada:2005bii} by adopting a family of three-body variational wavefunctions
 $|\Psi^{J^{PC}}_{ggg}\rangle$, and examining $\frac{\langle\Psi^{J^{PC}}|H^g_{\text{eff}}|\Psi^{J^{PC}}_{ggg}\rangle}{\langle\Psi^{J^{PC}}_{ggg}|\Psi^{JPC}_{ggg}\rangle}=M^{J^{PC}}_{ggg}$.
 
In $SU(3)$, the three-gluon ground state appears to be   
 $0^{-+}$, with a mass twice that of the scalar glueball, around $3900$ MeV in this model. But then the odd glueballs or oddballs can be predicted and again reasonable agreement with lattice theory is reached. 
 
In summary, the phenomenological Hamiltonian approaches inspired by QCD give reasonable few-body descriptions of the spectrum obtained via lattice methods and make the Yang-Mills mass gap part of the Clay Millenium Problem quite natural.
Now we are ready to address the original contribution of this article, the glueball-glueball interaction in the distance regime where the Van der Waals interaction is of interest. The overarching question we want to address is whether the glueball mass can be sizeably affected by the interaction (that is, the interaction energy is ever of the same order as the glueball mass in the GeV range).

\section{Matrix element for glueball-glueball interactions in generic terms}
\label{sec:VdW}

To the extent that a potential is a reasonable description, as the comparison with lattice data indicates, there can be a range of distances between glueballs at which the interaction is of the Van der Waals type. Its extraction from the underlying potential proceeds by analyzing the quantum theory of the London force.
Early work at Orsay~\cite{Gavela:1979zu} examined the residual interactions among conventional mesons and hadrons
($q\bar{q}$ and $qqq$) which we now extend to pure Yang-Mills states.

We may divide the Hamiltonian of the two-glueball system $|gg\rangle_1\otimes |gg\rangle_2$ as
$H=H_0+\mathcal{V}$: with kinetic term $H_0$ including the gluon dynamical mass justified in section~\ref{sec:salto de masa} which we will here take as constant, so that
 $H_0=\sum_{i=1}^4\sqrt{\boldsymbol{p}^2_i+m_g^2}$. On the other hand, $\mathcal{V}$ represents the interaction potential, typically
 \begin{align}
    \mathcal{V}=-\sum_{a<b}(V(\boldsymbol{r}_a-\boldsymbol{r_b})T_a^AT_b^A)
    \label{potglueb}
\end{align}
which displays Casimir scaling (so it cannot be the ultimate description of Confinement~\cite{Greensite2011}, but is useful in many instances and certainly at intermediate range).

The potential $V(\boldsymbol{r}_i-\boldsymbol{r_j})$ 
can be taken as the Coulomb one (at short range) or {\it e.g.}   the Cornell one.
The generators of the colour Lie group $T^A_a$ are in the adjoint representation: we use the uppercase superindex to identify each of the colour matrices and the lowercase subindices to denote the gluon pair over which it will act, in compact notation.\\

\subsection{Multipole expansion}

We perform a Feshbach reduction of the full Hamiltonian $H$ to the Hilbert subspace $P$ consisting of two colour-singlet glueballs (since we want to extract the interaction between them) with other states belonging to the subsystem $Q$. The reduced Hamiltonian can then be written as 
\begin{equation}
H_{\rm eff}=PHP+PHQ(E-QHQ)^{-1}QHP\ ,
\label{Feshbach}
\end{equation}
where $E=2E_1$ is the energy of the singlet glueball -singlet glueball state  $|11\rangle$, with $E_1$ that of an isolated glueball; $P=|11\rangle\langle11|$ projects over that state;  and  $Q=1-P=\sum |\chi_1\chi_2\rangle\langle\chi_1\chi_2|$  (summing over all colour configurations except the singlet-singlet one in $|11\rangle$, that is, over $|88\rangle$, $|88'\rangle$, etc. as explained below in Subsec.~\ref{subsec:2glueballcolor}) projects over the orthogonal complement in which the two glueballs are polarized, that is, not separately in colour-singlet states.

The effective Hamiltonian $H_{\rm eff}$ is constructed from the underlying interaction in $V_{\rm int}$ from Eq.~(\ref{Cornellpot}); the interglueball force therein makes sense for distances larger than the glueball size (internal scale). This is a good point to introduce some detailed geometric notation.
 
 First, let us define each glueball's centre of mass (cm),
 \begin{equation}
     \boldsymbol{R}_{(1)}=\frac{\boldsymbol{r}_1+\boldsymbol{r_2}}{2};\ \ \ \ \boldsymbol{R}_{(2)}=\frac{\boldsymbol{r}_3+\boldsymbol{r_4}}{2}\ ;
 \end{equation}
Then each gluon can be referred to the local-glueball cm,
\begin{equation}
    \boldsymbol{\rho}_a=\boldsymbol{r}_a-\boldsymbol{R}_{(1)};\ \ \ \ \boldsymbol{\rho}_b=\boldsymbol{r}_b-\boldsymbol{R}_{(2)}
    \label{GluonCoords}
\end{equation}
(where the subindices $a=1,2$ and $b=3,4$ run over the two gluons in each of the clusters)
and the glueball-glueball relative vector between the two clusters 
\begin{equation}
    \boldsymbol{R}=\boldsymbol{R}_{(1)}-\boldsymbol{R}_{(2)}\ .
\end{equation}
Therefore, the vector linking two gluons in different glueballs can be expressed as
\begin{equation}
    \boldsymbol{r}_a-\boldsymbol{r}_b=\boldsymbol{R}+(\boldsymbol{\rho}_a-\boldsymbol{\rho}_b)
\end{equation}

With this notation, the multipoles arise from a Taylor expansion of the interacting potential around 
$\boldsymbol{r}_a-\boldsymbol{r}_b=\boldsymbol{R}$, to second order:

\begin{align}
    V_{\rm int}(\boldsymbol{r}_a-\boldsymbol{r}_b)\simeq-\sum_{\substack{a,b}}T^A_aT^A_b \nonumber \\ \left(\! \!V(\boldsymbol{R})\!+\!(\rho_a^k-\rho_b^k)\left.\frac{\partial V}{\partial x^k}\right|_{\boldsymbol{R}}\!\!+\!(\rho_a^k-\rho_b^k)(\rho_a^l-\rho_b^l)\left.\!\frac{\partial^2V}{\partial x^k\partial x^l}\right|_{\boldsymbol{R}}\right)
    \nonumber \\
    \label{ec:taylor}
\end{align}

Equation (\ref{ec:taylor}) is best rewritten in terms of the colour-dipole moments, which for the individual gluons are 
\begin{equation}\boldsymbol{D}^A_a=\boldsymbol{\rho}_aT^A_a\ ,
\label{defdipolo}\end{equation} 
and for a composite glueball  read, combining those, 
\begin{equation}
\boldsymbol{D}^A_{(1)}=\boldsymbol{\rho}_1T^A_1+\boldsymbol{\rho}_2T^A_2;\ \ \ \ \boldsymbol{D}^A_{(2)}=\boldsymbol{\rho}_3T^A_3+\boldsymbol{\rho}_4T^A_4\ .
\label{GlueballDipole}
\end{equation}  

Thus, to zeroth order, the potential becomes
\begin{equation}
V^{(0)}=-T^A_{(1)}T^A_{(2)}V(\boldsymbol{R})\ ;
\end{equation} 
to first order,
\begin{equation}V^{(1)}=-(\boldsymbol{D}^A_{(1)}T^A_{(2)}-\boldsymbol{D}^A_{(2)}T^A_{(1)})\boldsymbol{\nabla}V(\boldsymbol{R})\ ; 
\end{equation}
and finally, to second order, 
\begin{equation}
V^{(2)}=-\frac{\partial^2V}{\partial x^k\partial x^l}(\frac{1}{2}\sum_{ab}T^A_aT^A_b(\rho^k_a\rho^l_a+\rho^k_b\rho^l_b)-(D^{A,k}_{(1)}D^{A,l}_{(2)}))\ .
\label{2ndordermultipole}
\end{equation}

\subsection{colour}\label{subsec:2glueballcolor}

Each two-gluon glueball belongs to the singlet representation contained in the decomposition
\begin{equation}
    8\otimes8=1\oplus8\oplus8'\oplus10\oplus\bar{10}\oplus27\ ,
\end{equation} 
but the whole basis of states to which the full Hamiltonian can connect is only required to be in an overall colour singlet.

Four gluons span $(8\otimes8)\otimes(8\otimes8)$ with dimension $8^4=4096$, but accepting only overall singlets, the combinations of the product representations
which can appear are
$1\otimes1$, $8\otimes8$, $8\otimes8'$, $8'\otimes8$, $8'\otimes8'$, $10\otimes\bar{10}$, $\bar{10}\otimes10$ and $27\otimes27$ so that the relevant Hilbert space is of dimension 8.

Decomposing the Hamiltonian into intracluster $H_i$ and intercluster $V_{\rm int}$ terms
within a nominal glueball or straddling the two, 
 $H=H_1+H_2+V_{\rm int}$, with $V_{\rm int}$ the potential of Eq.~(\ref{potglueb}), 
we proceed to the matrix elements of the interaction.

Easily, the matrix elements $\langle11|V_{\rm int}|11\rangle$ all vanish: this is because each term of $V^{(0)}$ and $V^{(1)}$ is proportional to $\langle1|T^A_{(1,2)}|1\rangle$ and $T^A_{(1,2)}$ is the adjoint generator of colour $SU(3)$ for either glueball, which is a colour singlet, since $1\not\subset 8\otimes 1$.

By the same token, the second order sum (first term of $V^{(2)}$ in Eq.~(\ref{2ndordermultipole}) also vanishes because the two relative-position operators are referred from a gluon to the glueball it belongs to, so that upon taking the matrix element, again a colour factor  $\langle1|T^A_{(1,2)}|1\rangle$ appears, 
with the same null value (a manifestation of Wigner-Eckart's theorem).

Finally, the diagonal colour-dipole contribution (``permanent dipole'' of each glueball) is also zero because of the colour factors 
$\langle 1|D^{A,k}_{(1,2)}|1\rangle$ and $  \langle 1| D^{A,l}_{(2,1)}|1\rangle$ which all vanish as they also contain an isolated $T^a$ each -see Eq.~(\ref{defdipolo}).

Therefore,  
\begin{eqnarray}
    \langle11|H|11\rangle&=&\langle11|H_1|11\rangle+\langle11|H_2|11\rangle\nonumber \\ &=&2E_1 
    PHP \nonumber \\
    &=&2E_1|11\rangle\langle11|
\end{eqnarray} 

Turning to the off-diagonal elements linking the two-singlet-glueball states to the rest of the linear space, 
\begin{eqnarray}
PHQ&=&\sum_{\chi\neq1}\langle11|H|\chi_1\chi_2\rangle|11\rangle\langle\chi_1\chi_2|
\nonumber \\
QHP&=&\sum_{\chi\neq1}\langle\chi_1\chi_2|H|11\rangle|\chi_1\chi_2\rangle\langle11| \ , 
\end{eqnarray}
we need the matrix elements
$\langle11|H|\chi_1\chi_2\rangle$ times $|\chi_1\chi_2\rangle$
which is one of the basis states distinct from $|11\rangle$.

A further simplification due to the colour algebra arises when applying the operator $T^A_{(i)}$ to the $i$th \emph{singlet} glueball, yielding zero: \begin{eqnarray}
T^A_{(i)}|1_{(i)}\rangle &=&  (T^A_1\otimes \mathbf{1}_2 +\mathbf{1}_1 \otimes T^A_2) (|a\rangle \times |b\rangle )\frac{\delta_{ab}}{\sqrt{8}} \nonumber \\
& =& -i\left(f^{Aa\lambda}|\lambda\rangle \otimes | b\rangle
+f^{Ab\lambda} |a\rangle\otimes |\lambda\rangle \right)\frac{\delta_{ab}}{\sqrt{8}}
\nonumber \\
\end{eqnarray}
which vanishes because of the antisymmetry of the structure constants $f$ (the matrix elements of $T$ in the adjoint representation).
This can be exploited as follows:
 \begin{align*}
     \langle11|V^{(0)}|\chi_1\chi_2\rangle=
     \\
     -V(\boldsymbol{R)}\langle1|T^A_{(1)}|\chi_1\rangle\langle1|T^A_{(2)}|\chi_2\rangle=0\\
     \langle11|V^{(1)}|\chi_1\chi_2\rangle=-\boldsymbol{\nabla}V(\boldsymbol{R})(\langle1|\boldsymbol{D}^A_{(1)}|\chi_1\rangle)\langle1|T^A_{(2)}|\chi_2\rangle \\ \qquad \qquad \ \   -\langle1|\boldsymbol{D}^A_{(2)}|\chi_2\rangle)\langle1|T^A_{(1)}|\chi_1\rangle)\\
      =-\boldsymbol{\nabla}V(\boldsymbol{R})(0-0)=0\\ 
    \langle11|V^{(2)}|\chi_1\chi_2\rangle = \\
    -\frac{\partial^2V}{\partial x^k\partial x^l} \Bigg[ \frac{1}{2}\sum_{a,b} \Big( \langle 1|T^A_a\rho^k_a\rho^l_a|\chi_1\rangle \langle 1|T^A_b|\chi_2\rangle \\
    \qquad + \langle 1|T^A_b\rho^k_b\rho^l_b|\chi_2\rangle \langle 1|T^A_b|\chi_1\rangle \Big)  \\
    \qquad  - \langle 1|D^{A,k}_{(1)}|\chi_1\rangle\langle1|D^{A,l}_{(2)}|\chi_2\rangle \Bigg]\\
    =\frac{\partial^2V}{\partial x^k\partial x^l}\langle 1|D^{A,k}_{(1)}|\chi_1\rangle\langle1|D^{A,l}_{(2)}|\chi_2\rangle
\end{align*}

It is also straightforward to see that
\begin{equation}
\langle11|H_1+H_2|\chi_1\chi_2\rangle=0\ ,
\end{equation} because $\chi_c$ 
is an eigenstate of $H_c$ for c=1,2 and the singlet ket is orthogonal to any other colour representation, $\langle1|\chi_c\rangle=0$. 

Therefore, with the help of the colour algebra, we have reduced
the off-diagonal matrix element between the two subspaces to
\begin{equation}
\langle11|H|\chi_1\chi_2\rangle=\frac{\partial^2V}{\partial x^k\partial x^l}\langle 1|D^{A,k}_{(1)}|\chi_1\rangle\langle1|D^{A,l}_{(2)}|\chi_2\rangle
\label{offdiagonalaftercolor}
\end{equation}
which we keep for later.

At last, we need the matrix elements on the nonstandard glueball Fock sector which is eliminated in the Feshbach procedure,
\begin{equation}QHQ=\sum_{\chi\neq1}\langle\chi_1\chi_2|H|\chi_1\chi_2\rangle|\chi_1\chi_2\rangle\langle\chi_1\chi_2|\ ;
\end{equation}
since we address the regime in which the interglueball distance is larger than the glueball size, we use the multipole expansion of Eq.~(\ref{ec:taylor}). We keep only the leading order in the expansion as this operator ends up in the denominator of Eq.~(\ref{Feshbach}). Thus, 
\begin{equation}
    \langle\chi_1\chi_2|V^{(0)}|\chi_1\chi_2\rangle=-V(\boldsymbol{R})\langle\chi_1\chi_2|T^A_{(1)}T^A_{(2)}|\chi_1\chi_2\rangle
\end{equation}. 

Now, $T^A_{(1)}$ and $T^A_{(2)}$ are the colour generators of each glueball, entailing that
$T^A_{(1)}+T^A_{(2)}$ is that for the full system, and since the final state
 $|\chi_1\chi_2\rangle$ is a colour singlet, we have
 \begin{equation}(T^A_{(1)}+T^A_{(2)})|\chi_1\chi_2\rangle=0\ .
 \end{equation}
 Hence, 
 \begin{eqnarray} 0 &=& 
 \langle\chi_1\chi_2| (T^A_{(1)}+T^A_{(2)})^2|\chi_1\chi_2\rangle \nonumber \\
 &=&\langle\chi_1|(T^A_{(1)})^2|\chi_1\rangle\langle\chi_2|\chi_2\rangle
 \nonumber \\
 & &+\langle\chi_2|(T^A_{(2)})^2|\chi_2\rangle\langle\chi_1|\chi_1\rangle
  \nonumber \\& & +2\langle\chi_1\chi_2|T^A_{(1)}T^A_{(2)}|\chi_1\chi_2\rangle\nonumber \\
  &=&C_2(\chi_1)+C_2(\chi_2)
   \nonumber \\
  & &+2\langle\chi_1\chi_2|T^A_{(1)}T^A_{(2)}|\chi_1\chi_2\rangle\ ,
 \end{eqnarray} 
 with $C_2(\chi_j)$ the quadratic Casimir of
 the indicated representation.

 In this way,
 \begin{eqnarray}
 \langle\chi_1\chi_2|V^{(0)}|\chi_1\chi_2\rangle &=& \frac{1}{2}(C_2(\chi_1)+C_2(\chi_2))V(\boldsymbol{R})\nonumber \\ &=& C_2(\chi_1)V(\boldsymbol{R})\ , 
 \end{eqnarray}
since the two representations $\chi_1,\chi_2$ 
combined to form an overall singlet are either equal or have equal quadratic Casimir.

Turning to the kinetic part of the Hamiltonian,
 \begin{equation}\langle\chi_1\chi_2|H|\chi_1\chi_2\rangle=2E_{\chi_1}+C_2(\chi_1)V(\boldsymbol{R}) \ ,
 \end{equation}
 and putting both together we arrive at
\begin{eqnarray}E-QHQ=2E_1|11\rangle\langle11|
\nonumber \\ +\sum_{\chi\neq1}(2E_1-(2E_\chi+C_2(R)V(\boldsymbol{R}))|\chi_1\chi_2\rangle\langle\chi_1\chi_2|)\ .
 \end{eqnarray} 

 We need to invert it for Eq.~(\ref{Feshbach})
but because the operator is diagonal, this is straightforward
\begin{eqnarray}
(E-QHQ)^{-1}=(2E_1)^{-1}|11\rangle\langle11|
\nonumber \\ +\sum_{\chi\neq1}((2E_1)-(2E_\chi+C_2(R)V(\boldsymbol{R})))^{-1}|\chi_1\chi_2\rangle\langle\chi_1\chi_2|)\ .
\nonumber \\
\end{eqnarray}

\subsection{Resulting Effective Hamiltonian}
Thanks to the large simplifications of the colour algebra, we have arrived at the following effective Hamiltonian for the two singlet glueballs under interaction,
 \begin{eqnarray}
H_{eff}=2E_1|11\rangle\langle11|
\nonumber \\ +\sum_{|\chi_1\chi_2\rangle\neq|11\rangle} \left( \frac{|\langle11|V_{\rm int}|\chi_1\chi_2\rangle|^2}{2E_1-(2E_{\chi}+C_2(\chi_1)V(\boldsymbol{R}))} \right)|11\rangle\langle11| \ . \nonumber \\
\label{Heff}
 \end{eqnarray}

As the most relevant example, taking the particular case of the Cornell potential $V(\boldsymbol{r})= a r-\frac{\alpha_s}{r}$ for the underlying colour interaction in Eq.~(\ref{Heff}) yields an effective potential
of the following form:
 \begin{eqnarray}
     V_{eff}(R)=\nonumber \\  \sum_{|\chi_1\chi_2\rangle\neq|11\rangle}
          \frac{1}{2E_1-(2E_{\chi}+C_2(\chi)(aR-\alpha_sR^{-1}))}\times \nonumber \\
     \left( 
     (\frac{a}{R}+\frac{\alpha_s}{R^3})\langle11|\boldsymbol{D}^A_{(1)}\cdot\boldsymbol{D}^A_{(2)}|\chi_1\chi_2\rangle \right.
      \nonumber \\  \left.
     -(\frac{a}{R^3}+\frac{3\alpha_s}{R^5})\langle1|x_kD^{A,k}_{(1)}|\chi_1\rangle\langle1|x_lD^{A,l}_{(2)}|\chi_2\rangle
    \right)^2.\
     \nonumber \\ 
     \label{Poteff}
 \end{eqnarray}

 \section{Longer and shorter range contributions with the Cornell potential}
 The evaluation of this potential in Eq.~(\ref{Poteff}) is not straightforward as it requires knowledge, not only of the singlet-glueball wavefunction on the brac, which is relatively well-known~\cite{Llanes-Estrada:2000ozq}, but also that of higher representations $\chi_j$ on the ket, which make sense only within the larger cluster to achieve a global colour singlet. This requires solving a four-body problem where knowledge is more scarce~\cite{Boulanger:2008aj}
particularly with regard to decomposing the states in two-body--two-body clusters in various colour representations. We will therefore postpone a full numerical evaluation  of the result in Eq.~(\ref{ec:asintVdW}) to future work and content ourselves with getting an idea of its size for a Coulombic system, and later providing a very schematic treatment of the linear part in appendix~\ref{app:includelinear}. Since the gluon mass gap is of order 0.8 GeV, which is not that far from the charm quark mass at 1.2 GeV, solidly capturing the short-range part of the potential, at least, should bear some resemblance with the full calculation.

\subsection{$1/R^3$ longer-range tail}
Before turning to the short-range piece, let us also quickly give the large-distance behaviour of Eq.~(\ref{Poteff}). This is easily expressed upon writing the numerator in terms of the tensor
\begin{equation}\label{projector}
    P_{kl}:=\frac{\delta_{kl}-x_kx_l/R^2}{R}
\end{equation} as
\begin{equation}
a^2P_{kl}P_{ij}\langle11|D_{(1)}^{Ak}D_{(2)}^{Al}|\chi_1\chi_2\rangle\langle11|D_{(1)}^{Ai}D_{(2)}^{Aj}|\chi_1\chi_2\rangle\ .
\end{equation}

Upon summing over all glueball polarizations in the kets (the nonsinglet $\chi_j$ colour representations)
$|\chi_1,m_1\rangle$ and $|\chi_2 m_2\rangle$,
and using rotational invariance,
\begin{eqnarray}\label{sumoverpols}
\sum_{m_1,m_2}\langle11|D_{(1)}^{Ak}D_{(2)}^{Al}|\chi_1\chi_2\rangle\langle11|D_{(1)}^{Ai}D_{(2)}^{Aj}|\chi_1\chi_2\rangle=
\nonumber \\ \frac{1}{3}\delta^{ki}\delta^{lj}|\langle11|\boldsymbol{D}_{(1)}^A\cdot\boldsymbol{D}^A_{(2)}|\chi_1\chi_2\rangle|^2\ . 
\end{eqnarray}
The polarization tensor $\delta^{ki}\delta^{lj}$ then contracts the operators in the numerator as $P_{kl}P^{kl}=2/R^2$. 

In turn, the denominator, at large distance, is simply $C_2(\chi)aR$. 

Therefore, we find what we believe to be a new observation for the glueball-glueball interaction for large (but not asymptotically large) $R$, that the potential of Eq.~(\ref{Poteff}) behaves as
 \begin{align}
     V_{\rm eff}(R)\rightarrow-\frac{2a}{3R^3}\sum_{\chi\neq1}\frac{|\langle11|\boldsymbol{D}^A_{(1)}\cdot\boldsymbol{D}^A_{(2)}|\chi_1\chi_2\rangle|^2}{C_2(\chi)}
     \label{ec:asintVdW}
 \end{align}
 decreasing with the power $R^{-3}=R^{1-4}$, 
 (subtracting 4 from the dominant power of the colour potential, just as in the known quark-antiquark meson case~\cite{Gavela:1979zu}).

\subsection{$1/R^6$ shorter-range behaviour}
We now turn to the direct analogue of the traditional London force in molecular physics, the shorter-range part from the $1/r$ underlying Coulomb potential piece.  This is dominant when $\alpha_s/R>aR$ or below about 0.3 fm with the parameters employed: at this short distance glueballs are already overlapping so that it is outside the validity window of the Van der Waals approach. But the computation serves as template for the full Cornell potential.

Dropping this time the terms in the numerator of Eq.~(\ref{Poteff}) which involve the linear potential and momentarily employing the letters $A$, $B$ for its two terms, the numerator of Eq.~(\ref{Poteff}) reads
\begin{eqnarray}
{\rm Numerator}&=& \frac{\alpha_s^2}{R^4}\left(\frac{A}{R}+\frac{B}{R^3}\right)^2 ;
\nonumber \\
    A&\equiv& \langle11|\boldsymbol{D}^A_{(1)}\!\! \cdot\!\!\boldsymbol{D}^A_{(2)}|\chi_1\chi_2\rangle\\
    B&\equiv&-\frac{3x_kx_l}{R^2}\langle1|D^{A,k}_{(1)}|\chi_1\rangle\langle1|D^{A,l}_{(2)}|\chi_2\rangle
    \nonumber \\
   {\rm Numerator} &=&
\frac{\alpha_s^2}{R^4}Q_{ki}Q_{ij}
\langle 11|D^{A,k}_{(1)}D^{A,a}_{(2)}
|\chi_1\chi_2\rangle \nonumber\\
&&{}\times
\langle 11|D^{A,i}_{(1)}D^{A,j}_{(2)}
|\chi_1\chi_2\rangle .
\end{eqnarray}
with  $Q_{kl}$ the following tensor
\begin{equation} \label{skewedprojector}
    Q_{kl}=\frac{\delta_{kl}-3x_kx_l/R^2}{R}.\
\end{equation}
Upon summing over all glueball polarizations $m_j$ (excepting the colour singlet's, of course), $|\chi_1,m_1\rangle,|\chi_2 m_2\rangle$ 
due to invariance under rotations,
\begin{eqnarray}
    \sum_{m_1,m_2}\langle11|D_{(1)}^{Ak}D_{(2)}^{Al}|\chi_1\chi_2\rangle\langle11|D_{(1)}^{Ai}D_{(2)}^{Aj}|\chi_1\chi_2\rangle\nonumber \\
    =\frac{1}{3}\delta^{ki}\delta^{lj}|\langle11|\boldsymbol{D}_{(1)}^A\cdot\boldsymbol{D}^A_{(2)}|\chi_1\chi_2\rangle|^2\ ,
\end{eqnarray}
and using $Q_{kl}Q^{kl}=6/R^2$ the numerator of Eq.~(\ref{Poteff})  becomes
\begin{eqnarray}
{\rm Numerator} &=& 
        \frac{\alpha_s^2}{3R^4}\delta^{ki}\delta^{lj}Q_{kl}Q_{ij}|\langle11|\boldsymbol{D}_{(1)}^A\cdot\boldsymbol{D}^A_{(2)}|\chi_1\chi_2\rangle|^2 \nonumber \\&=& 
    \frac{\alpha_s^2}{3R^4}Q_{kl}Q^{kl}|\langle11|\boldsymbol{D}_{(1)}^A\cdot\boldsymbol{D}^A_{(2)}|\chi_1\chi_2\rangle|^2 \nonumber \\ &=& 
    \frac{2\alpha_s^2}{R^6}|\langle11|\boldsymbol{D}_{(1)}^A\cdot\boldsymbol{D}^A_{(2)}|\chi_1\chi_2\rangle|^2
\end{eqnarray}

The denominator is very simple, $ 2E_1-(2E_\chi-\alpha_s/R)$
and thus, in the regime where the Coulomb piece exceeds the linear one, the effective potential turns out to be
    \begin{align}
        \boxed{V_{\rm eff}(R)=\frac{2\alpha_s^2}{R^6} \sum_{|\chi_1\chi_2\rangle\neq|11\rangle}\frac{|\langle11|\boldsymbol{D}_{(1)}^A\cdot\boldsymbol{D}^A_{(2)}|\chi_1\chi_2\rangle|^2}{2E_1-(2E_\chi-C_2(\chi)\frac{\alpha_s}{R})}}
        \label{London_Coulomb}
    \end{align}
which we will numerically evaluate.
 In Eq.~(\ref{Potentialperchannel})  and~(\ref{AntisymmetricOctet}), given in Appendix~\ref{app:wfsCoulomb}, we will show that with the $SU(3)$ colour algebra, the only representation in the sum over $|\chi_1\chi_2\rangle$ that will make a contribution is the antisymmetric octet, and the rest actually yield zero.

\section{Coulomb-potential evaluation}
The task at hand is to calculate the square moduli of the
$\langle11|\boldsymbol{D}_{(1)}^A\cdot\boldsymbol{D}^A_{(2)}|\chi_1\chi_2\rangle$  dipole matrix elements

Because the system is two-body and Coulombic, the wavefunctions are hydrogenlike (see Appendix~\ref{app:wfsCoulomb}) and the hydrogen atom selection rules likewise apply.
    
For a dipole transition, $\Delta l=\pm 1$, so that if we follow the Van der Waals force among ground state scalar glueballs ($l=0$), the octet representation in $\chi$ must be an  $l=1$ P-wave glueball. Thus, we only need $R_{n1}(r;a)$. But there is no particular selection rule restricting the radial excitation of the polarized glueball, so that we will work with a generic $n$ and sum over this quantum number.

The radial-coordinate part of each matrix element requires the following integral,
\begin{equation} \label{RadialIntDef}
    I_n:=\langle 10|r|n1\rangle=\int_0^\infty r^3 R_{1,0}(r,a_1)R_{n,1}(r,a_8)dr
\end{equation}
which is handled in Appendix~\ref{app:wfsCoulomb}.

Returning to $|\langle11|\boldsymbol{D}_{(1)}^A\cdot\boldsymbol{D}^A_{(2)}|\chi_1\chi_2\rangle|^2$, 
we need matrix elements of the colour-dipole operator in each glueball, $\langle1|D^{A,i}_{(a)}|\chi\rangle$. 
Decomposing the colour dipole of each glueball using Eq.~(\ref{GlueballDipole}) and the gluon coordinates of Eq.~(\ref{GluonCoords}) we rewrite the dipole-dipole operator as

\begin{eqnarray}
\boldsymbol{D}_{(1)}^A\cdot\boldsymbol{D}^A_{(2)}&=&\boldsymbol{\rho}_1\cdot\boldsymbol{\rho}_3T^A_1T^A_3+\boldsymbol{\rho}_1\cdot\boldsymbol{\rho}_4T^A_1T^A_4 \nonumber \\
& + & 
\boldsymbol{\rho}_2\cdot\boldsymbol{\rho}_3T^A_2T^A_3+\boldsymbol{\rho}_2\cdot\boldsymbol{\rho}_4T^A_2T^A_4
\label{DipoleDipoleCoulomb}
\end{eqnarray}
from which we will only need the dipole matrix elements
 $\langle11|\boldsymbol{D}_{(1)}^A\cdot\boldsymbol{D}^A_{(2)}|\chi_1\chi_2\rangle$ for the antisymmetric octet representation,  $|\chi_i\rangle = |8_A\rangle$, $i=1,2$ as the decuplet and 27-plet are shown in Appendix~\ref{app:wfsCoulomb} to have nonattractive potentials, and the symmetric octet has zero matrix element because of gluon-exchange symmetry.

With this in mind, the numerator of Eq.~(\ref{London_Coulomb}) now reads 
\begin{widetext}
\begin{eqnarray} \label{DipolesinglettoAoctet}
    \langle 11|\boldsymbol{D}_{(1)}^A\cdot\boldsymbol{D}^A_{(2)}|8_A 8_A\rangle&=\frac{1}{\sqrt{8}}\frac{9}{24}8\langle100;100|
\boldsymbol{\rho}_1\cdot\boldsymbol{\rho}_3-\boldsymbol{\rho}_1\cdot\boldsymbol{\rho}_4-\boldsymbol{\rho}_2\cdot\boldsymbol{\rho}_3+\boldsymbol{\rho}_2\cdot\boldsymbol{\rho}_4|n_11m_1;n_21m_2\rangle\\
&=\frac{3}{\sqrt{8}}\langle100;100|
\boldsymbol{\rho}_1\cdot\boldsymbol{\rho}_3-\boldsymbol{\rho}_1\cdot\boldsymbol{\rho}_4-\boldsymbol{\rho}_2\cdot\boldsymbol{\rho}_3+\boldsymbol{\rho}_2\cdot\boldsymbol{\rho}_4|n_11m_1;n_21m_2\rangle
\end{eqnarray}
\end{widetext}

But because 
\begin{equation}
    \boldsymbol{\rho}_1+\boldsymbol{\rho}_2=\boldsymbol{r}_1+\boldsymbol{r}_2-2\boldsymbol{R}_{(1)}=0\ ,
\end{equation}
(and the same for 3 and 4) we have that 
\begin{equation}
\boldsymbol{\rho}_2=-\boldsymbol{\rho}_1\ \ \ \ \boldsymbol{\rho}_4=-\boldsymbol{\rho}_3
\end{equation}
which greatly simplifies Eq.~(\ref{DipolesinglettoAoctet}) to
\begin{eqnarray}
    \langle 11|\boldsymbol{D}_{(1)}^A\! \! \cdot\!\! \boldsymbol{D}^A_{(2)}|8_A 8_A\rangle
&=\frac{12}{\sqrt{8}}\langle100;100|
\boldsymbol{\rho}_1\cdot\boldsymbol{\rho}_3|n_11m_1;n_21m_2\rangle
\ . \nonumber  \\
\end{eqnarray}
Remembering the internal relative coordinates of each of the two glueballs, $\boldsymbol{r}_{(1)}=\boldsymbol{r}_1-\boldsymbol{r}_2$ and $\boldsymbol{r}_{(2)}=\boldsymbol{r}_3-\boldsymbol{r}_4$ 
the scalar products can be rendered as
\begin{equation} \label{DipoleSingletAoctet2}
    \boldsymbol{\rho}_1\cdot\boldsymbol{\rho}_3=\frac{1}{4}(\boldsymbol{r}_1-\boldsymbol{r}_2)\cdot(\boldsymbol{r}_3-\boldsymbol{r}_4)=\frac{1}{4}\boldsymbol{r}_{(1)}\cdot\boldsymbol{r}_{(2)}
\end{equation} 
so we get an angular dependence on the relative orientation of the two glueballs,
\begin{equation}
\boldsymbol{r}_{(1)}\cdot\boldsymbol{r}_{(2)}=r_{(1)}r_{(2)}cos(\gamma)\ .
\end{equation}
Substituting back into Eq.~(\ref{DipoleSingletAoctet2}) we recognize the need for the evaluation of angular integrals which are presented in Appendix~\ref{app:angular}.

From that angular reduction, we get an effective Van der Waals potential depending only on the radial integrals defined in Eq~(\ref{RadialIntDef}),
\begin{align}
    V_{\rm eff}(R)=\frac{9\alpha_s^2}{4R^6} \sum_{n_1,n_2}\frac{I_{n_1}^2I_{n_2}^2}{2E_1-E_{8,n_1}-E_{8,n_2}+\frac{3\alpha_s}{R}}\ .
\end{align}

The sum extends over all radially-excited, $P$-wave, \\antisymmetric-octet glueballs, characterized by the integers 
 $n_1$ and $n_2$.
 
 The denominator has been slightly generalized as we do not assume here that the two excited glueballs have equal energy as in $2E_8=E_8+E_8$ (they do not need to be the first excitation anymore). Instead, the energy of a glueball in the $\chi$ representation with principal quantum number $n$ will be
\begin{align}
    \boxed{E_{\chi,n}=-\frac{\mu}{2n^2}\alpha_{\rm eff}^2(\chi)}
\end{align}
Particular cases of interest here are
\begin{equation} 
E_1=E_{1,1}=-\frac{9}{2}\mu\alpha_s^2\ ,   \ \ \  E_{8,n}=-\frac{9}{4}\mu\frac{\alpha_s^2}{n^2}\ .
\end{equation}

Thus, for the Van der Waals potential derived from an underlying colour-Coulomb potential among gluons we have a closed expression,
\begin{align}
    \boxed{V_{\rm eff}(R)=-\frac{3\alpha_s}{4R^6} \sum_{n_1,n_2}\frac{I_{n_1}^2I_{n_2}^2}{3\mu\alpha_s(1-(2n_1)^{-2}-(2n_2)^{-2})-R^{-1}}}\ .
    \label{ec:pot ef coulomb}
\end{align}

Remembering now that the radial integrals $I_n$ really depend on the coupling $\alpha_s$ implicitly through the Bohr radius $a_1$, 
that  dependence can be extracted thanks to their linearity in this variable,
\begin{align}
    I_n(a_1)=a_1I_{n,0}
\end{align}
where $I_{n,0}=I_n(a_1=1)$ 
is the adimensional radial integral obtained by setting $a_1=1$ as indicated. 

Taking into account that
$a_1=2/(3m_g\alpha_s)$ we see that
\begin{align}
    V_{\rm eff}(R)=-\frac{4\alpha_s^{-3}}{27R^6} \sum_{n_1,n_2}\frac{I_{n_1,0}^2I_{n_2,0}^2}{3\mu\alpha_s(1-(2n_1)^{-2}-(2n_2)^{-2})-R^{-1}}\ .
\end{align}
Rather counterintuitively, this potential is more binding the smaller $\alpha_s$ is!

The way to understand it is to think that the average glueball radius in this Coulombic approximation
\begin{equation}
   \langle r\rangle=\frac{3}{2}a_1\propto\frac{1}{\alpha_s} 
\end{equation}
increases with decreasing coupling constant. But then the polarizability~\cite{Atkins} can be larger because the charges are more separated, so the effect of the radii overcomes that of the coupling constant.
(This of course breaks down as $\alpha_s\to 0$ when the glueball radius $\langle \rho \rangle $ diverges and the Van der Waals potential has no domain of applicability $R>2\langle \rho \rangle$.)

\subsection{Comparison with the Yukawa/Dilaton Glueball-Glueball potential}\label{subsec:dilaton}

To have a reference potential we turn to the Yukawa interaction due to glueball exchange. As argued in subsection~\ref{subsec:incompatibilidadVdW}, this must be the dominant potential among two glueballs at asymptotically large distances, but how relevant it is at intermediate distances is to be ascertained. 

The Yukawa potential is of course of the following form (and universally attractive sincee the lightest Yang-Mills glueball is scalar): 
\begin{equation} \label{Yukawapotential}
V(r) = -\frac{g^2}{4\pi} \frac{e^{-m_G r}}{r}
\ .
\end{equation}
Naturally, the $m_G$ parameter is the lightest glueball mass (in $SU(3)$ with the QCD scale, about 1.7 GeV).
The coupling however is less obvious. 

A possibility is to adopt a dilaton-type~\cite{Migdal:1982jp} effective Lagrangian $\mathcal{L}_{\rm eff}(G)$ for the scalar (composite-particle) glueball field $G$, which additionally reproduces the trace anomaly,

\begin{eqnarray} \label{traceanomaly}
\partial_\mu J^\mu_{\rm dilatation} = T_\mu^\mu &=& 4V-G\frac{\partial V}{\partial G}\nonumber \\
&\underset{=}{?}& \frac{-11N_c}{24}\langle \frac{\alpha_s}{\pi}
F^a_{\mu\nu}F^{a\ \mu\nu}\rangle\ .
\end{eqnarray}
Taking a lattice-inspired size~\cite{Giacosa:2021brl} of the gluon-condensate scale as $C\simeq 0.55$GeV
(with the expectation value in Eq.~(\ref{traceanomaly}) being
$C^4\equiv  \langle \frac{\alpha_s}{\pi}
F^a_{\mu\nu}F^{a\ \mu\nu}\rangle$)
and matching the loss of scale invariance of the spontaneous breaking in the effective Lagrangian to the anomalous breaking of the fundamental theory yields the following dilaton model with coupling parameter $\Lambda_G\simeq 0.5$ GeV for $N_c=3$,
\begin{eqnarray}
    \mathcal{L}_{\rm eff}(G) &=& \frac{1}{2}(\partial_\mu G\partial^\mu G) -\frac{1}{4} \frac{m_G^2}{\Lambda_G^2}\left(
    G^4\log\left| \frac{G}{\Lambda_G}\right| -\frac{G^4}{4} \right)
    \nonumber \\
\end{eqnarray}
(which yields $T^\mu_\mu = 4V-G\partial_G V = -\frac{1}{4}\frac{m_G^2}{\Lambda^2}G^4$, or at the minimum, $\frac{-1}{4} m_G^2\Lambda_G^2$ from which $\Lambda_G$ is obtained upon comparing with the Yang-Mills anomaly).
The first two terms of the series expansion of the potential term in the Lagrangian in powers of $G$ are a vacuum constant and a mass term,
\begin{equation}
\mathcal{V}^{0,2}(G) ={\rm constant} + \frac{1}{2}m_G^2 G^2 \ .
\end{equation}
The third-order term
\begin{equation}
\mathcal{V}^{3}(G) =  \left(5\frac{m_G^2}{\Lambda_G^2}\right) \frac{G^3}{3!}
\end{equation}
provides, upon comparing with the standard Feynman vertex of scalar theory, $-g\ G^3/3!$, the coefficient needed for Eq.~(\ref{Yukawapotential}) and thus a model Yukawa potential for the glueball-to-glueball coordinate ${\bf R}$,
\begin{equation}\label{Yukawapotential2}
V(R) = -\frac{25(m_G/\Lambda_G)^4}{4\pi} \frac{e^{-m_GR}}{R}\ .
\end{equation}

\begin{figure}
    \centering
    \includegraphics[width=0.9\linewidth]{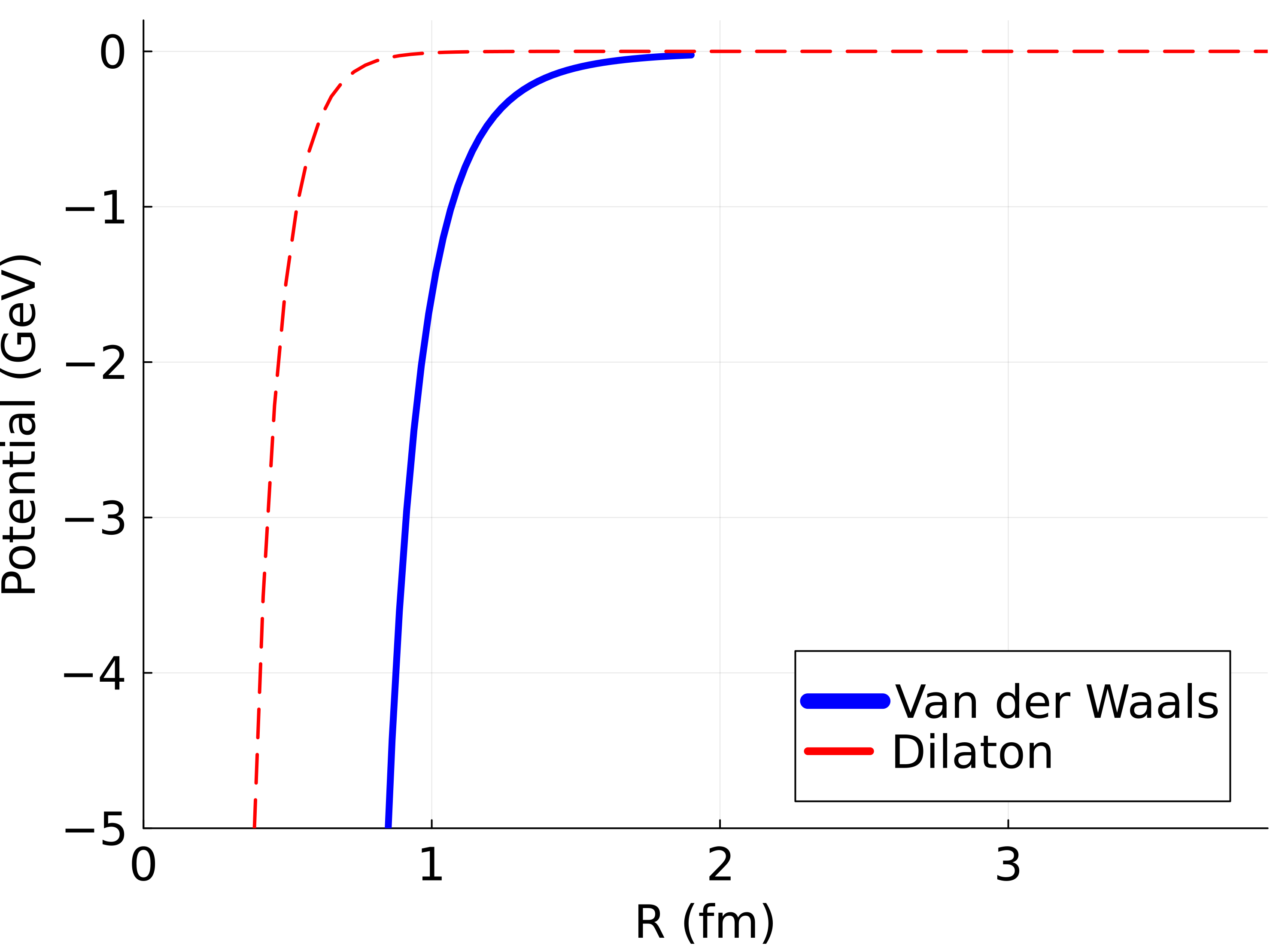}
    \caption{ Comparison of the glueball-mediated Yukawa potential of Eq.~(\ref{Yukawapotential2})
    with the Van der Waals potential of Eq.~(\ref{ec:pot ef coulomb}) in its regime of validity (only the contribution from the colour Coulomb potential is included, for an estimate with the linear piece see Appendix~\ref{app:includelinear}).  
    Over most of its range of validity $(0.66,1.9)$ fm (that is, between the two glueballs at contact and  the string breaking length affecting the underlying potential between polarized octet glueballs in the intermediate state)
    the Coulomb Van der Waals potential is seen to be dominant. But this impression changes upon examining the results of section~\ref{sec:fullpotential}. 
    \label{fig:comparison}}
\end{figure}

This is to be compared with the Van der Waals potential which we have been developing, culminating in Eq.~(\ref{ec:pot ef coulomb}), comparison shown in Fig.~\ref{fig:comparison}.

Finally, we note that the tree-level contribution to the scattering amplitude of the next term in the Taylor expansion of $\mathcal{V}(G)$, namely
\begin{equation}
\mathcal{V}^4(G)=\left(11\frac{m_G^2}{\Lambda_G^2}\right) \frac{G^4}{4!}
\end{equation}
yields a repulsive contact potential
\begin{equation}
V_{\rm contact}({\bf R}) = \left(11\frac{m_G^2}{\Lambda_G^2}\right)
\delta^{(3)}({\bf R})\ .
\end{equation}
Because it is a positive contribution, it weakens even more the one-glueball exchange potential of Eq.(\ref{Yukawapotential2}), so it does not invalidate our reasoning.

\section{Numerical computation with the complete Cornell potential}\label{sec:fullpotential}

We now undertake a numerical solution of the radial  wave equation in full. Given that it has no natural zero-energy point, the Cornell potential of Eq.~(\ref{Cornellpot}) then calls for an additional constant $C$
\begin{equation}
V_{\rm int}(|\boldsymbol{x}-\boldsymbol{y}|)= C -\frac{\alpha_s}{|\boldsymbol{x}-\boldsymbol{y}|}+ a|\boldsymbol{x}-\boldsymbol{y}|\ ,
\label{Cornellpot2}
\end{equation}
to be normalized to the ground state glueball mass, which we set from lattice gauge theory, rounding off, at 1.7 GeV. 
 Neither the radial eigenfunctions nor the Van der Waals potential are modified by this change, but it does affect the spectrum and is necessary in a quantum-mechanical setup (not so in a field-theory treatment). 

Reabsorbing it in the masses, the radial equation reads

\begin{equation}
H_\chi =
-\frac{1}{2\mu}\frac{d^2}{dr^2}
+K_\chi\left(ar-\frac{\alpha_s}{r}\right)
+\frac{l(l+1)}{2\mu r^2},
\label{eq: Hamchi}
\end{equation}
with $K_\chi=\left(C_A-C_2(\chi)/{2}\right)$.
For $SU(3)$, the singlet and octet colour factors are
\begin{equation}
K_1=3,\qquad K_8=\frac{3}{2}\ .
\end{equation}
We therefore obtain the singlet ground state from
\begin{equation}
H_1 =
-\frac{1}{2\mu}\frac{d^2}{dr^2}
+3\left(ar-\frac{\alpha_s}{r}\right),
\end{equation}
and the octet intermediate states from
\begin{equation}
H_8 =
-\frac{1}{2\mu}\frac{d^2}{dr^2}
+\frac{3}{2}\left(ar-\frac{\alpha_s}{r}\right)
+\frac{l(l+1)}{2\mu r^2},
\qquad l=1\ .
\end{equation}
Again, the octet states are supposed to be out of the spectrum due to colour confinement, and are to be understood as part of an overall singlet with more components, here another such octet glueball.
Solving the radial equation yields the singlet ground state $E_1^{(1)}$ and a tower of excited octet $E_8^{(n)}$ energies.

With the eigenfunctions for the hamiltonian from Eq. (\ref{eq: Hamchi}) at hand, we can now extract the mean value of the radius of the ground-state singlet scalar glueball,  $\langle r\rangle$. We simply carry out the following integral
\begin{equation}
    \langle r\rangle=\int_0^\infty r|u_1^{(1)}(r)|^2dr,
\end{equation}
with $u_1^{(1)}$  having unit square norm, to obtain
\begin{eqnarray}
    \langle r\rangle_{SU(3)}\approx 0.316\rm{fm}\nonumber \\
    \langle r\rangle_{SU(2)}\approx 0.386\rm{fm}
\label{radiuscalc}
\end{eqnarray}
for the $SU(3)$ and $SU(2)$ groups respectively, and the same values of $\alpha_s$ and string tension $a$ (the one in $SU(2)$ should be marginally smaller but we ignore the difference). 

It is reassuring to note that the glueball radius value for $SU(3)$ is very close, within $5\%$, to the 0.33 fm obtained by Shuryak and Zahed \cite{Shuryak:2026grt},
 by completely different means.

With some confidence in the eigenfunctions, the radial
overlap integrals are numerically evaluated  as
\begin{equation}
I_n=
\int_0^\infty dr\,
u_1^{(1)}(r)\,r\,u_8^{(n)}(r).
\end{equation}

The angular and colour dipole matrix element 
has the form of Eq.~(\ref{dipolematrixelement})
except that the Coulomb wavefunctions are substituted by the numerically computed ones for the full Cornell potential.


Consequently, the Van der Waals potential of Eq.~(\ref{eq: Poteffgeneral}) becomes
\begin{eqnarray}  
 V_{\rm eff}(R)=\frac{3}{8}\left(\frac{2a^2}{R^2}+\frac{6\alpha_s^2}{R^6}+\frac{4a\alpha_s}{R^4}\right) \times
    \nonumber \\
\sum_{n,m}
\frac{I_n^2I_m^2}
{
2E_1^{(1)}
-E_8^{(n)}
-E_8^{(m)}
-3\left(aR-\frac{\alpha_s}{R}\right)
}.\nonumber\\
\label{eq: VdWSU(3)numericocompleto}
\end{eqnarray}

The sum is truncated to the first $N$ octet states and its
convergence is checked by varying $N$.\\
We can also render this potential for the $SU(2)$ group (to facilitate comparison with Lattice Gauge Theory), by effecting  minor changes in the numerical factors only,
\begin{eqnarray}
V_{\rm eff}^{SU(2)}(R)
=
\frac{4}{9}
\left(
\frac{2a^2}{R^2}
+\frac{6\alpha_s^2}{R^6}
+\frac{4a\alpha_s}{R^4}
\right)\times
\nonumber\\
\sum_{n,m}\frac{I_n^2I_m^2}
{
2E_1^{(1)}
-E_3^{(n)}
-E_3^{(m)}
-2\left(aR-\frac{\alpha_s}{R}\right)
}.\nonumber\\
\label{eq:VdWnumericoSU(2)}
\end{eqnarray}

\begin{figure}
    \centering
    \includegraphics[width=\linewidth]{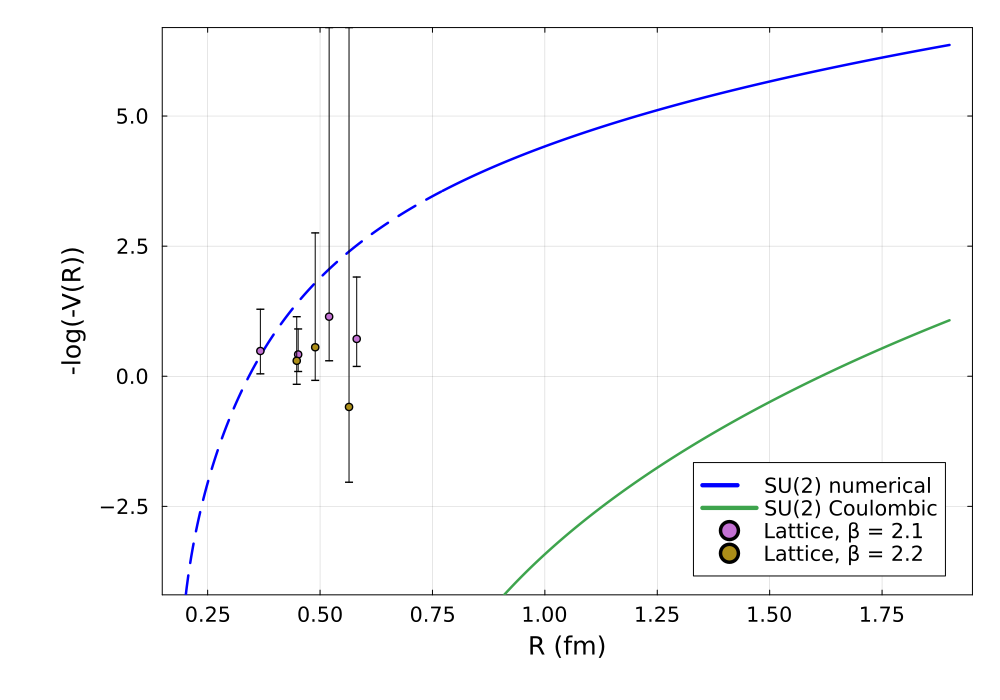}
    \caption{
     Glueball potential in GeV, in a logarithmic scale. (decreases towards the top and right).\\
    Pure Coulomb- (lower line, solid, way too large respect to the lattice data); full Cornell-induced Van der Waals potential, which is smaller but saturates the extant lattice data~\cite{Yamanaka:2019aeq} (symbols) which has been computed on a regime where the two glueballs overlap, so one should not speak of glueball-glueball potential anyway (which is why we dash that line).
    The data extraction employs Eq. (3) of~\cite{Yamanaka:2019aeq},  to convert the scale $\Lambda$ to $0.27 \rm{GeV}$ (fixed by the $SU(2)$ glueball mass) using the value of $\sigma=0.18\rm{GeV}^2$.}
        \label{fig: comparacionSU(2)}
\end{figure}

In Figure (\ref{fig: comparacionSU(2)}) we compare the numerical Eq. (\ref{eq:VdWnumericoSU(2)}) (derived from the Cornell potential) and the analytical Eq. (\ref{eq:VdWSU(2)Coulomb}) (derived from the Coulomb potential) Van der Waals potentials for the group $SU(2)$ with the lattice calculations.

Noting from Eq.~(\ref{radiuscalc}) that $2\langle r\rangle_{SU(2)}\approx0.77\rm{fm}$, which should be the lowest distance between glueballs where a potential can be extracted, we find surprising that lattice calculations \cite{Yamanaka:2019aeq} have been performed for $r<0.4\Lambda^{-1}$; if we use their conversion (for $\beta=2.5$) to fm, we obtain a value of $\approx 0.6\rm{fm}$, well below the 0.77 fm at which the average radii overlap in $SU(2)$. So it would seem that some lattice calculations are performed in a region where the two glueballs are sitting on top of each other, and it does not make too much sense to speak of a potential among them. 

Still, we see that the Van der Waals potential extracted from the Cornell potential saturates the lattice data easily.

\begin{figure}
    \centering
    \includegraphics[width=0.9\linewidth]{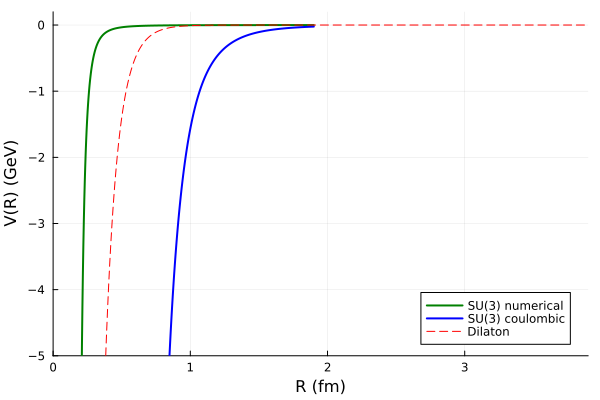}
    \caption{ We compare the Van der Waals potential for purely Coulombic interactions among constituent gluons and that obtained with a full Cornell potential (also inserted is the dilaton model for reference).
    The different sign of the two contributions in 
    Eq. (\ref{eq: VdWSU(3)numericocompleto}) leads to very important cancellations for the ground states.
    This means that the overall mass gap is barely affected by glueball-glueball attraction.
    (We should note that this cancellation might be less pronounced for excited-glueball interactions, and also remind the reader that the region of validity of this potential is the interval $(0.66,1.9)\rm{fm}$.)}
    \label{fig:ComparacionSU(3)}
\end{figure}

In Figure (\ref{fig:ComparacionSU(3)}) we compare Van der Waals potentials (for the group $SU(3)$) from Eq. (\ref{eq: VdWSU(3)numericocompleto}) and Eq. (\ref{ec:pot ef coulomb}) with the potential obtained from the dilaton model Eq. (\ref{Yukawapotential2}).

That dilaton effective field theory model yields too large an interaction, but the authors of that computation~\cite{Giacosa:2021brl} acknowledge that they have an adjustable constant which may lower the potential sufficiently (they were not in possession of this lattice data back then). 

\section{Cross section in s-channel}\label{sec:cross}
Of further interest for investigations into dark sectors, and perhaps for relativistic heavy-ion collisions is glueball-glueball scattering.
The quantity of primary concern is the low-energy cross-section, expressible in terms of the $s$-wave phase shift
\begin{equation}
    \sigma_0(k)=\frac{4\pi}{k^2}\sin^2(\delta_0(k))
    \label{eq:sigma0}
\end{equation}
which dominates the partial-wave series for small $k$. 

We are in a position to calculate it from the Van der Waals potential. A sleek way to do it is solving Calogero's phase equation
\begin{equation}
    \frac{d\delta_0(k,R)}{dR}=-\frac{2\mu}{k}V_{\rm eff}(R)\sin^2(kR+\delta_0(k,R))\ .
    \label{eq: Calogero}
\end{equation}
which is an ordinary, first-order but nonlinear differential equation in $R$ for an auxiliary function $\delta_0(k,R)$  ($k$ being the scattering momentum). This function grows by accruing the contribution of a potential slice at distance $R$ from the center of force, to yield, once the potential has decayed enough (in our case, 1.9 fm where one should transition to the exponentially suppressed Yukawa potential), the phase shift  $\delta_0(k)=\delta_0(k,R_{\rm end})$ .

We integrate it starting at the point of contact around 0.66 fm where the phase shift is taken to vanish 
\begin{equation}
    \delta_0(k,R=2\langle r\rangle)=0\ ,
\end{equation}
employing 
the Van der Waals potential $V_{\rm eff}(R)$ numerically computed for the underlying Cornell potential, from Eq.~(\ref{eq: VdWSU(3)numericocompleto}), setting $\mu$ (the reduced mass of the glueball-glueball system) to $\mu=M_G/2\approx m_g$.

We solve the equation twice for each of the
$SU(3)$ and $SU(2)$ groups. Again, $\alpha_s$ and $a$ are taken the same in the two cases, but the colour factors make a difference.

\begin{figure}
    \centering
    \includegraphics[width=0.95\linewidth]{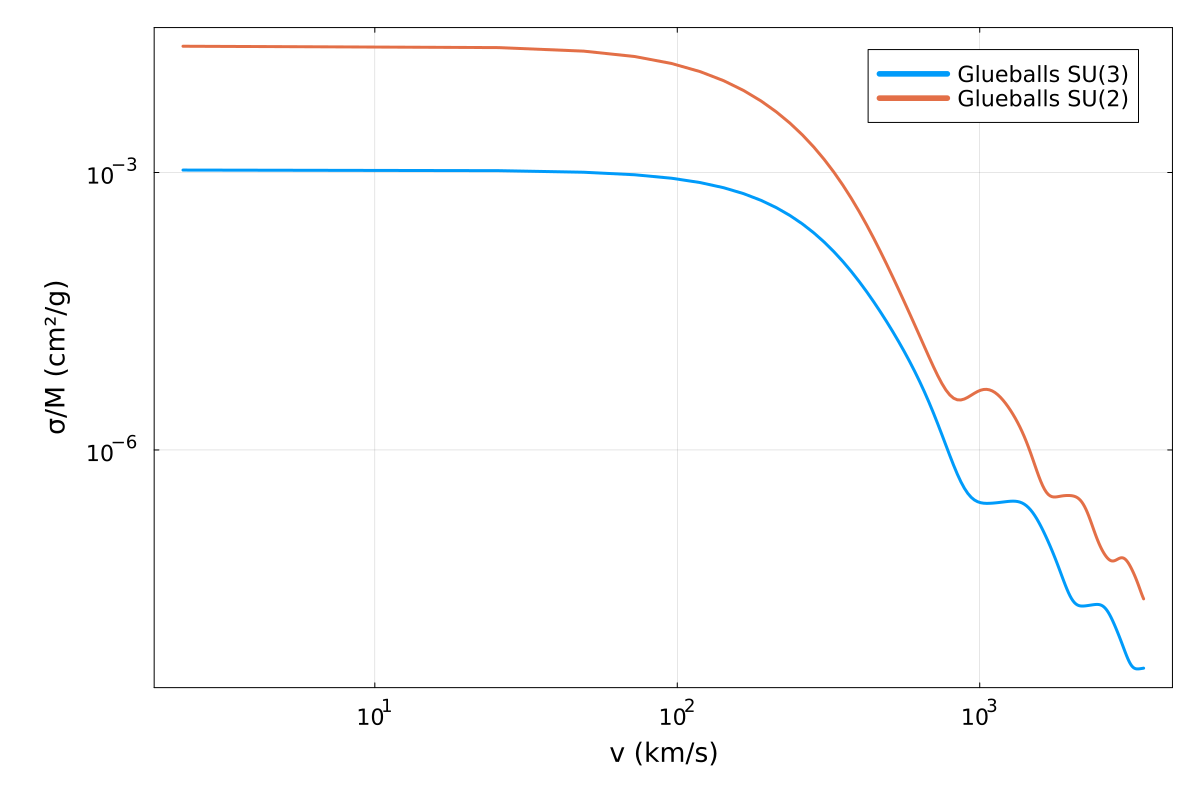}
    \caption{Velocity-dependence of the partial-wave cross section $\sigma_0(k)$ for $SU(3)$ (top, red online) and $SU(2)$ (bottom, blue online). }
    \label{fig:crosssection}
\end{figure}

We obtain 
\begin{eqnarray}
\sigma^{SU(3)}(0)\approx3.22\rm{mb} \nonumber \\
\sigma^{SU(2)}(0)\approx66.48\rm{mb}\ . 
\label{crosssections}
\end{eqnarray}
An effect contributing to a larger $SU(2)$ cross section is the larger glueball radius.
It is difficult to state how  these cross sections vary with the  parameters $\sigma$, $\alpha_s$ because of the intricate dependence through wavefunctions and also through the spectrum, $M_G$ (and thus $\mu$, which is indirectly determined). This calls for quite some additional numerical analysis which exceeds this paper. We do provide the dependence with $k$ of the $s$-wave cross section in Fig.~\ref{fig:crosssection}.

In dark matter studies it is usual to refer cross sections to the unit mass, here the glueball one $M_G$,
$\sigma_0(0)/M_G$ which we quote next,
\begin{eqnarray}
\frac{\sigma^{SU(3)}_{l=0}(k=0)}{M_G}&=&1.06\times10^{-3}\ \rm{cm}^2/\rm{g}
\nonumber \\
 \frac{\sigma^{SU(2)}_{l=0}(k=0)}{M_G}&=& 2.31\times10^{-2}\ \rm{cm}^2/\rm{g}\ .
\label{crossSecperunitM}
\end{eqnarray}
In those, the glueball masses have been calculated with the same parameter set for both groups, $a=0.18\rm{GeV}^2$ and 
$\alpha_s=0.4$ for consistency with the rest of the article. They result in $M_G^{SU(3)}=1.7\ \rm{GeV}$ (which is in agreement with $M_G(0^{++})=3.8\sqrt{a}$ from lattice calculations \cite{MichaelTeper1987}) and $M_G^{SU(2)}=1.61\ \rm{GeV}$.

The $SU(3)$ cross section at least seems to be too small to be of astrophysical use (this number in this units would have to be of order 0.1 to have an effect on cosmological simulations of the core/cusp issue), but of course, there is a broad parameter space: we only take of it that a precise copy of Yang-Mills theory with our physical parameters does not seem to be viable as a dark sector.

\section{Conclusions}

In this article about pure Yang-Mills glueballs we have obtained a closed analytical expression for the effective induced-dipole to induced-dipole colour potential with an underlying Cornell (funnel) potential.  We have also pursued a full evaluation for the Coulomb part thereof. 

A further evaluation with the entire Cornell potential would require additional numerical work, although recently an interesting analytical representation has been reported~ \cite{Napsuciale:2025rdr}.

As usual for the London force, the interaction falls off with a sixth power of the distance between glueballs (a third power in case the underlying linear potential is relevant).

We have found that the structure of the colour group through the values of its Casimir invariants restricts the possible intermediate states which can be polarized from the ground state scalar and singlet glueball, to basically the antisymmetric-octet glueball: the symmetric octet, the decuplets and 27plet  do not contribute to the Van der Waals interaction.

This makes a numerical evaluation, based on Eqs.~(\ref{ec:pot ef coulomb}), (\ref{ec: N_n}) and (\ref{ec: I_n}) quite straightforward.
This must be considered as very schematic since precise numerical values require further analysis showing the sensitivity to variations of the scale $\mu$ controlling $\alpha_s$, and especially the gluon mass gap $m_g$ on which the calculation does have nonnegligible dependence.

Temptatively, we find that the London force among scalar ground-state glueballs is  attractive. Its Coulomb potential contribution is large but is rather compensated by the opposing linear piece to yield a more moderate  glueball-glueball interaction. This makes it sensitive to parameter choices such as the interaction scale, on which $\alpha_s$ depends. Still, the Van der Waals potential contribution  corresponding to the linear piece in Eq.~(\ref{ec:asintVdW})  in spite  of its different sign with respect to the Coulomb one in Eq.~(\ref{London_Coulomb}), does not alter our conclusion.

Of course, further spin-dependent potentials can be active at shorter range (they typically have increasing powers of $1/r$) and relativistic effects need to be taken into account in further work. But perhaps a first task for that future would be to try to develop an effective field theory, at the level of the glueballs, which reproduces the Van der Waals interaction
at intermediate ranges. 
This is, to our knowledge, absent from the literature, although generally known as WEFT~\cite{Brambilla:2017ffe}, in which the Van der Waals potential appears as a matching coefficient of the Effective Field Theory. 
\medskip 

Given that 
\begin{itemize}
\item Glueballs are colour singlets;
\item Glueballs are composite and colour-polarizable;
\item And there is a significant Yang-Mills mass gap suppressing Yukawa exchanges, which provide the full interaction only at large distances, where they are very small;
\end{itemize}
We think that our conclusion is inescapable, independently of model details to realize the evaluation of the dipole matrix elements, and that a Van der Waals potential needs to be taken into account in any treatment of glueball scattering.

\section*{Acknowledgments}

FJLE acknowledges partial support from grants PID2025-170350NB-I00 and PID2025-169724NB-C21 of the spanish government's program MICIU/AEI/10.13039/501100011033 and of ERDF/EU.

\clearpage
\appendix

\section{Coulomb wavefunctions and radial integrals} \label{app:wfsCoulomb}
\subsection{Colour factor for the internal glueball potential}

We need the glueball wavefunctions in various colour representations, in the approximation in which the gluon-gluon interaction is Coulombic, to compute at least that contribution to the Van der Waals force.
Therefore the reduced-particle in the two-gluon system is subject to a colour-dependent potential
    \begin{equation}
        V(r)=\frac{\alpha_s}{r}T^A_1T^A_2
    \end{equation}
    with $T^A_i$ the colour generator for the  $i$th gluon of the two; the colour generator corresponding to the entire virtual glueball (which is here not a singlet, but in the $\chi$ representation) is $T^A=T^A_1+T^A_2$. This yields~\cite{Slansky:1981yr}
    \begin{eqnarray}
        C_2(\chi)=T^AT^A&=&(\underbrace{T_1^AT_1^A}_{C_2({\rm octet})=3}+\underbrace{T_2^AT_2^A}_{C_2({\rm octet})=3}+2T_1^AT_2^A)
        \nonumber \\
        T_1^AT_2^A&=&\frac{1}{2}C_2(\chi)-3
    \end{eqnarray}
    Within a generic representation of $SU(3)$ characterized by Dynkin indices  $(p,q)$, the Casimir operator can be given as the identity times a numerical coefficient~\cite{Slansky:1981yr}
    \begin{align*}
        C_2(p,q)=\frac{1}{3}(p^2+q^2+pq+3p+3q)
    \end{align*}
   which for convenience is tabulated in table~\ref{tab:casimir} for the particular representations here needed.

In consequence, the potential among the gluons within the same glueball in each $\chi$ representation is
\begin{eqnarray}
    V(r,\chi)&=&-\frac{\alpha_s}{r}\left(3-\frac{C_2(\chi)}{2}\right) \nonumber \\
    V(r,1)&=& -\frac{3\alpha_s}{r} \nonumber \\
    V(r,8)&=&-\frac{3\alpha_s}{2r} \nonumber \\
    V(r,10)&=&0   \nonumber \\
    V(r,27)&=&\frac{\alpha_s}{r} \ .
    \label{Potentialperchannel}
\end{eqnarray}
Obviously, the only two representations admitting Coulomb bound states are the singlet and the octet. Decuplet and antidecuplet are degenerate and ``free'' (in this approximation), and the 27-plet features a repulsive potential.

We can employ the shorthand
 $\alpha_{eff}(\chi)=\alpha_s(3-C_2(\chi)/2)$
 and then the Bohr radius for a bound state (which exists when $\alpha_{eff}(\chi)>0$) is
\begin{align}
    a_{\chi}=\frac{1}{\mu\alpha_{eff}(\chi)}=\frac{2}{m_g\alpha_{eff}(\chi)}
\end{align}
where $\mu=m_g/2$ is the reduced effective gluon mass. 
The singlet and octet representations, in particular, have the following Bohr radii
\begin{eqnarray}
    a_{1}=\frac{2}{3m_g\alpha_s} \nonumber \\
    a_{8}=\frac{4}{3m_g\alpha_s}=2a_{1}
\end{eqnarray}

\begin{table}
\centering
\begin{tabular}{|c|c|c|}
\hline
\textbf{Representation} & \textbf{$(p,q)$} & \textbf{$C_2$} \\ \hline
$8$ & $(1,1)$ & $3$ \\ \hline
$10$ & $(3,0)$ & $6$ \\ \hline
$\overline{10}$ & $(0,3)$ & $6$ \\ \hline
$27$ & $(2,2)$ & $8$ \\ \hline
\end{tabular}
\caption{\label{tab:casimir}  Casimir coefficients for the virtual-glueball representations
(in the understanding that the intermediate states will always have two such glueballs 
so as to yield an overall colour-singlet state). 
}
\end{table}
\subsection{Radial integrals}
Having characterized the necessary Coulomb potentials we may examine now their eigenstates,
which we factorize into space and colour (and polarization $\lambda$, omitted hereafter) wavefunctions. 
\begin{equation}|\chi;\lambda\rangle=\underbrace{|\rm{c}_\chi;\lambda\rangle}_{\rm{colour}}\otimes |nlm\rangle
\end{equation}
Ssince the singlet glueball is by hypothesis in the ground state, its wavefunction will be
$$|1\rangle=\underbrace{|\rm{c_1\rangle}}_{\rm{colour}}\otimes |100\rangle$$
The spatial part solves a Hydrogen problem with coupling constant depending on the representation $\chi$, and is of the form
\begin{equation}
    \psi_{nml}^{(\chi)}(r,\theta,\varphi)=R_{n,l}(r,a_\chi)Y_{l,m}(\theta,\varphi)
\end{equation}

The radial part acquires its colour-representation dependence through the Bohr radius,
\begin{eqnarray}
    R_{nl}(r;a)&=&
\left(\frac{2}{na}\right)^{3/2}
\sqrt{\frac{(n-l-1)!}{2n(n+l)!}}\nonumber \\
& & \times e^{-r/(na)}
\left(\frac{2r}{na}\right)^l
L_{n-l-1}^{\,2l+1}\!\left(\frac{2r}{na}\right)
\end{eqnarray}
which for the singlet ground state is the usual
\begin{align}
    \psi^{(1)}_{100}(r)&=\frac{1}{\sqrt{\pi a_1^3}}e^{-r/a_1} & R_{1,0}(r,a_1)=\frac{2}{\sqrt{a_1^3}}e^{-r/a_1}\ .
\end{align}

With them we can address the computation of the radial integrals in Eq.~(\ref{RadialIntDef}). The constants are grouped in a factor
\begin{equation}
    N_n(a_8)=\left(\frac{2}{na_8}\right)^{5/2}\sqrt{\frac{(n-2)!}{2n(n+1)!}}
    \label{ec: N_n}
\end{equation}
leaving the following expression for the  integrand of $I_n$,
\begin{equation}
    N_n(a_8)\frac{2}{\sqrt{a_1^3}}r^4e^{-\frac{2n+1}{2na_1}r}L^3_{n-2}\left(\frac{2r}{na_8}\right)\ .
\end{equation}
Because the Laguerre polynomials admit a closed expression as a finite sum,
\begin{align}
    L_n^\alpha(x)=\sum_{i=0}^n(-1)^i\begin{pmatrix}
        n+\alpha\\
        n-i
    \end{pmatrix}\frac{x^i}{i!} \ ,
\end{align}
the radial integral reduces to 
\begin{eqnarray} \label{radintegralintermediate}
    I_n=N_n(a_8)\frac{2}{\sqrt{a_1^3}}\sum_{i=0}^{n-2}
    \nonumber \\ 
    \frac{(-1)^i}{(na_1)^ii!}\begin{pmatrix}
        n+3\\
        n-i
    \end{pmatrix}  \int_0^\infty\left(r^{4+i}e^{-\frac{2n+1}{2na_1}r}\right)dr \ .
\end{eqnarray}
The integral is elementary. With the change of variables $u=\frac{2n+1}{2na_1}r$, 
\begin{eqnarray}
    \int_0^\infty\left(r^{4+i}e^{-\frac{2n+1}{2na_1}r}\right)dr &=& \nonumber \\ \left(\frac{2na_1}{2n+1}\right)^{5+i}\int_0^\infty\left(u^{4+i}e^{-u}\right)du&=& \nonumber \\\left(\frac{2na_1}{2n+1}\right)^{5+i}\Gamma(5+i)
\end{eqnarray}
which, when taken to Eq.~(\ref{radintegralintermediate}), gives 
\begin{align}
     I_n=N_n(a_8){2a_1^{7/2}}\sum_{i=0}^{n-2}
     \nonumber \\ 
     \frac{1 }{(-n)^i}
     \frac{(4+i)!}{i!}
     \begin{pmatrix}
        n+3\\
        n-i
    \end{pmatrix}\left(\frac{2n}{2n+1}\right)^{5+i}\ ,
    \label{ec: I_n}
\end{align}
a closed, analytical form for the radial integral entering
the $1\to n$ dipole-transition matrix element.
Having this series representation permits proceeding to very high values of $n$ without needing to numerically integrate over products of Laguere polynomials which, due to the increasing number of Sturm-Liouville zeroes as $n$ increases, becomes rather costly.

\subsection{Colour factors for the wavefunctions}

In this subappendix we engage the colour algebra necessary for Eq.~(\ref{DipoleDipoleCoulomb}).
Remember first that the glueball-glueball state $|\chi_1\chi_2\rangle$ is an overall colour singlet although each glueball separately might not be. 

Because we only need to work with the symmetric ($S$) or antisymmetric ($A$) octets. the global state must take the form
\begin{equation}
|c_{8_{(S/A)}}c_{8_{(S/A)}}\rangle=\frac{1}{\sqrt{8}}\delta^{ab}|c_{8_{(S/A)}};a\rangle\otimes |c_{8_{(S/A)}};b\rangle
\end{equation}
(where $a$, $b$ colour indices are contracted).

Each of the three representations which are still in the calculation at this point have the following glueball wavefunction in terms of the two constituent gluons:
\begin{eqnarray}
    |c_1\rangle&=&N_1\delta^{ab}|a\rangle\otimes |b\rangle \nonumber \\
    |c_{8A},c\rangle&=&N_{8A}f^{abc}|a\rangle\otimes |b\rangle \nonumber \\
    |c_{8S},c\rangle&=&N_{8S}d^{abc}|a\rangle\otimes |b\rangle
\end{eqnarray}
in terms of the $SU(3)$ structure constants $f^{abc}$ and its completely symmetric tensor $d^{abc}$. We quickly determine the normalization constants $N_1,N_{8A}$ and $N_{8S}$: 
\begin{eqnarray}
    1=\langle c_1|c_1\rangle&=&N_1^2\delta^{ab}\delta^{cd}\langle c|a\rangle\langle d|b\rangle\nonumber \\
    &=&N_1^2\delta^{ab}\delta^{cd}\delta_{ac}\delta_{bd}\nonumber  \\
    &=&8N_1^2 \implies \boxed{N_1=1/\sqrt{8}}\ ;
\end{eqnarray}
\begin{eqnarray}
    \delta^{cd}=\langle c_{8A},d|c_{8A},c\rangle&=&|N_{8A}|^2f^{abc}f^{pqd}\langle p|a\rangle\langle q|b\rangle \nonumber \\
    &=&|N_{8A}|^2f^{abc}f_{abd}\nonumber  \\
    &=&3|N_{8A}|^2\delta^{cd} 
    \nonumber \\
    &\implies& \boxed{N_{8A}=1/\sqrt{3}}\ ;
\end{eqnarray}
\begin{eqnarray}
    \delta^{cd}=\langle c_{8S},d|c_{8S},c\rangle&=&|N_{8S}|^2d^{abc}d^{pqd}\langle p|a\rangle\langle q|b\rangle\nonumber \\
    &=&|N_{8S}|^2d^{abc}d_{abd}\nonumber \\
    &=&\frac{5}{3}|N_{8S}|^2\delta^{cd} \nonumber \\
    &\implies& \boxed{N_{8S}=\sqrt{3/5}} \ .
\end{eqnarray}

Each term summed in Eq.~(\ref{DipoleDipoleCoulomb}) carries a product of two colour operators over a pair of gluons chosen from different glueballs, $T^A_aT^A_b$ with $a=1,2$ and $b=3,4$.
The colour part factorizes, and each matrix element is similar to, for example: 
\begin{eqnarray}
\langle c_1c_1|T^A_1T^A_3|c_{8_{(S/A)}};\lambda,c_{8_{(S/A)}};\mu\rangle= \nonumber \\ \langle c_1|T^A_1|c_{8_{(S/A)}};\lambda\rangle\langle c_1|T^A_3|c_{8_{(S/A)}};\mu\rangle \ .
\end{eqnarray}
Let us explicitly compute the first of the two factors in this last equation, for the symmetric octet,

\begin{eqnarray}
    \langle c_1|T^A_1|c_{8_A};\lambda\rangle&=&\frac{1}{\sqrt{24}}\delta^{ab}f^{cd\lambda}\langle ab|T^A_1| cd\rangle\nonumber \\
    &=&\frac{1}{\sqrt{24}}\delta^{ab}f^{cd\lambda}(-i)f^{Aec}\langle a|e\rangle\langle b|d\rangle\nonumber\\
    &=&\frac{1}{\sqrt{24}}\delta^{ab}f^{cd\lambda}(-i)f^{Aec}\delta_{ae}\delta_{bd}\nonumber \\
    &=&\frac{1}{\sqrt{24}}f^{ca\lambda}(-i)f^{Aac}\nonumber \\
    &=&\frac{i}{\sqrt{24}}C_A\delta^{\lambda A}=\frac{3i}{\sqrt{24}}\delta^{\lambda A}\ .
\end{eqnarray}

Then, thanks to the wavefunction symmetry, the other matrix elements within the glueball corresponding to each gluon easily fall off, $\langle c_1|T^A_1|c_{8_A};\lambda\rangle=-\langle c_1|T^A_2|c_{8_A};\lambda\rangle=\langle c_1|T^A_3|c_{8_A};\lambda\rangle=-\langle c_1|T^A_4|c_{8_A};\lambda\rangle$. 

But now, the symmetric octet simplifies to zero, 
\begin{eqnarray}
    \langle c_1|T^A_1|c_{8_S};\lambda\rangle&=&\sqrt{\frac{3}{40}}\delta^{ab}d^{cd\lambda}\langle ab|T^A_1| cd\rangle \nonumber \\
    &=&\sqrt{\frac{3}{40}}\delta^{ab}d^{cd\lambda}(-i)f^{Aec}\langle a|e\rangle\langle b|d\rangle\nonumber \\
    &=&\sqrt{\frac{3}{40}}\delta^{ab}d^{cd\lambda}(-i)f^{Aec}\delta_{ae}\delta_{bd} \nonumber\\
   &=&\sqrt{\frac{3}{40}}d^{ca\lambda}(-i)f^{Aac}=0
   \label{AntisymmetricOctet}
\end{eqnarray}
(the wavefunctions are gluon-permutation even, while the operator is odd, its matrix elements being the structure constants).

Therefore, and keeping in mind the result of Eq.~(\ref{Potentialperchannel}) where only singlets and octets led to binding, 
we conclude that {\it the only representation to which the singlet glueball virtually polarizes to produce a London force is the antisymmetric octet}

\section{Angular integration}\label{app:angular}

We now show the (standard) evaluation of the angular integral
in Eq.~(\ref{DipoleSingletAoctet2}).
For unit orbital angular momentum,  $P_1(cos(\gamma))=cos(\gamma)$, 
the addition theorem for spherical harmonic functions 
\begin{equation}
    P_l(cos(\gamma))=\frac{4\pi}{2l+1}\sum_{m=-l}^lY_{l,m}(\Omega_1)Y^*_{l,m}(\Omega_2)
    \end{equation}
directly gives
\begin{equation}
   \cos(\gamma)=\frac{4\pi}{3}\sum_{m=-1}^1Y_{1,m}(\Omega_1)Y^*_{1,m}(\Omega_2)
\end{equation}
Hence,
\begin{widetext}
\begin{align*}
    \langle 11|\boldsymbol{D}_{(1)}^A\cdot\boldsymbol{D}^A_{(2)}|8_A m_1;8_A m_2\rangle
&=\frac{3}{\sqrt{8}}\frac{4\pi}{3}\sum_{m=-1}^1\langle10|r_{(1)}|n_11\rangle\langle10|r_{(2)}|n_21\rangle\langle00;00|Y_{1,m}(\Omega_1) Y_{1,m}^*(\Omega_2)|1m_1;1m_2\rangle\\
&=\frac{3}{\sqrt{8}}I_{n_1}I_{n_2}\frac{4\pi}{3}\sum_{m=-1}^1\langle 00|Y_{1,m}(\Omega_1)|1m_1\rangle\langle00|Y^*_{1,m}(\Omega_2)|1m_2\rangle\ .
\end{align*}
The second line is the angular integration:
\begin{eqnarray} \label{angularintegration}
    \frac{4\pi}{3}\sum_{m=-1}^1\langle 00|Y_{1,m}(\Omega_1)|1m_1\rangle\langle00|Y^*_{1,m}(\Omega_2)|1m_2\rangle&=&\frac{4\pi}{3}\sum_{m=-1}^1\int Y_{00}^*(\Omega_1)Y_{1m}(\Omega_1)Y_{1m_1}(\Omega_1)d\Omega_1\times \nonumber \\& & \times \int Y_{00}^*(\Omega_2)Y_{1m}(\Omega_2)Y_{1m_2}(\Omega_2)d\Omega_2 \ .
\end{eqnarray}
\end{widetext}
Such angular integrals are given by the formula
\begin{eqnarray}
\int
Y_{l_1m_1}(\Omega)\,
Y_{l_2m_2}(\Omega)\,
Y_{l_3m_3}(\Omega)\,
d\Omega
= \nonumber \\
\sqrt{\frac{(2l_1+1)(2l_2+1)(2l_3+1)}{4\pi}}
\begin{pmatrix}
l_1 & l_2 & l_3\\
0 & 0 & 0
\end{pmatrix}
\begin{pmatrix}
l_1 & l_2 & l_3\\
m_1 & m_2 & m_3
\end{pmatrix}  \nonumber \\
\end{eqnarray}
but in this case they are particularly simple since $Y_{0,0}$ is a constant which can be extracted from the integral and leaves an orthogonality relation between two spherical harmonics.
The two integrals are therefore
\begin{eqnarray}
    \int Y_{00}^*(\Omega_1)Y_{1m}(\Omega_1)Y_{1m_1}(\Omega_1)d\Omega_1&=\frac{(-1)^m}{\sqrt{4\pi}}\delta_{m,-m_1}\nonumber \\
    \int Y_{00}^*(\Omega_2)Y_{1m}(\Omega_2)Y_{1m_2}(\Omega_2)d\Omega_2&=\frac{(-1)^m}{\sqrt{4\pi}}\delta_{m,-m_2}\ . \nonumber \\
\end{eqnarray}
Substituting back into Eq.~(\ref{angularintegration}) one obtains
\begin{align}
    \frac{4\pi}{3}\! \! \sum_{m=-1} ^1\! \langle 00|Y_{1,m}(\Omega_1)|1m_1\rangle\langle00|Y^*_{1,m}(\Omega_2)|1m_2\rangle=\frac{1}{3}\delta_{m_1,-m_2}
\end{align}
that is, we retrieve the already known factor of  $1/3$ coming from averaging on polarizations, which must not be double counted. 
In rigour, if we summed  the squared matrix elements over all the polarizations, we would get
\begin{equation}
    \sum_{m_1,m_2}\left(\frac{1}{3}\delta_{m_1,-m_2}\right)^2=3\frac{1}{9}=\frac{1}{3}
\end{equation}
as already advanced in Eq.~(\ref{sumoverpols}).

We have finally reduced the angular part of the computation and can write down the dipole-dipole operator in terms of the two radial integrals over Coulomb wavefunctions treated in Appendix~\ref{app:wfsCoulomb}, to wit
\begin{align}
  \label{dipolematrixelement}  \boxed{\langle 11|\boldsymbol{D}_{(1)}^A\cdot\boldsymbol{D}^A_{(2)}|8_A,n_1;8_A,n_2\rangle=\frac{3}{\sqrt{8}}I_{n_1}I_{n_2}} \ .
\end{align}
Equipped with this result we may return to Eq.~(\ref{DipoleSingletAoctet2}).

\section{Convergence of the series representation of the effective potential}
We now dedicate a few paragraphs to the convergence of Eq.~(\ref{ec:pot ef coulomb}).
Because the dipole excites the $l=0$ ground-state glueball to $l=1$, the summation over the principal quantum number starts with $n_1,n_2=2$.

The following auxiliary shorthand notation for the denominator will be useful,
\begin{align}
    D(n_1,n_2,R)=3\mu\alpha_s\left(1-\frac{1}{4n_1^2}-\frac{1}{4n_2^2}\right)-\frac{1}{R}\ .
\end{align}

The Feshbach Hilbert-space reduction and further treatment which we have employed necessitates that the denominator does not vanish, so that we may assume that $D(n_1,n_2,R)$ is of constant sign (with the example parameter set chosen, this sign change happens slightly beyond 2 fm). Calling $\delta$ the (positive) infimum of $D$ in the interval of interest,
\begin{equation}
  \frac{I_{n_1}^2I_{n_2}^2}{D(n_1,n_2,R)}\le\frac{I_{n_1}^2I_{n_2}^2}{\delta}\ .
\end{equation}
Then, the uniform convergence of the series of Eq.~(\ref{ec:pot ef coulomb})
 will fall off Weierstrass's $M$ criterion provided that 
\begin{align}
    \sum_{n_1=2}^\infty\sum_{n_2=2}^\infty I_{n_1}^2I_{n_2}^2
\end{align}
with $I_n=\langle 10|r|n1\rangle$
is convergent itself irrespective of the denominator.

To show that this is the case, let us focus on
\begin{align}
    \sum_{n=2}^\infty I_n^2=\sum_{n=2}^\infty|\langle 10|r|n1\rangle|^2 \ ;
\end{align}
the projector therein $P_p=\sum_n|n1\rangle\langle n1|$, over the subspace with $l=1$, must satisfy
\begin{align}
    \langle\psi|P_p|\psi\rangle\le\langle\psi |\psi\rangle=||\psi||^2
\end{align}
Thus, 
\begin{eqnarray}
     \sum_n^\infty I_n^2 &=& \langle 10|rP_pr|10\rangle
\nonumber \\     
     &\le& \langle10|r^2|10\rangle=3a_1^2<\infty\ ,
\end{eqnarray}
certainly convergent.

This guarantees certain properties:
\begin{itemize}
    \item The potential is a continuous function of $R$.
    \item Limits in $R$ and sum signs can be interchanged.
    \item The truncation of the series has a well defined bound on the error which we address shortly.
    \item The derivative of each term in the series may be separately taken.
\end{itemize}
Moreover, in addition to the uniform convergence, an upper bound on the potential can be found since the series with a simplified denominator
\begin{equation}
\sum _{n_1,n_2}I_{n_1}^2I_{n_2}^2/\delta\le (3a_1^2)^2/\delta
\end{equation}
is an upper bound to the series giving the potential,
\begin{align}
    V_{\rm eff}(R)\ge-\frac{3\alpha_s}{4}\frac{9a_1^4}{\delta R^6}=-\frac{4}{3m_g^4\delta}\ \frac{\alpha_s^{-3}}{R^6}
\end{align}

\subsection{Bounding the error upon truncating the series}
A numerical computation of the potential in   Eq.~(\ref{ec:pot ef coulomb}) 
requires retaining a finite number of terms only up to, say, $N$:
\begin{align}
    V_{{\rm eff},N}(R)=-\frac{3\alpha_s}{4R^6} \sum_{n_1,n_2}^N\frac{I_{n_1}^2I_{n_2}^2}{3\mu\alpha_s(1-(2n_1)^{-2}-(2n_2)^{-2})-R^{-1}} \ .
\end{align}
We now dedicate a few lines to constraining the truncation error incurred, namely
\begin{align}
    \mathcal{R}_N(R):=V_{\rm eff}(R)-V_{{\rm eff},N}(R)
\end{align}

To bound the error from above we shall use 
\begin{align}
    |\mathcal{R}_N(R)|\le \frac{3\alpha_s}{4R^6}\sum_{n_1>N\text{ ó }n_2>N}M_{n_1,n_2}
\end{align}
thanks to uniform convergence which provides such inequality when 
 $\sum_{n_1,n_2}M_{n_1,n_2}$ 
 is a series bounding 
the potential of Eq.~(\ref{ec:pot ef coulomb}) term by term.

Indeed, as we just saw, $M_{n_1,n_2}=I_{n_1}^2I_{n_2}^2/\delta$ satisfies that very property, and this leads to the inequality
\begin{eqnarray}
    |\mathcal{R}_N(R)|&\le & \frac{3\alpha_s}{4\delta R^6}\sum_{n_1>N\text{ or }n_2>N}I_{n_1}^2I_{n_2}^2
    \nonumber \\ 
    &\le& \frac{3\alpha_s}{4\delta R^6}\left(\sum_{n_1>N,n_2}I_{n_1}^2I_{n_2}^2+\sum_{n_2>N,n_1}I_{n_1}^2I_{n_2}^2\right)
        \nonumber \\ 
    &\le&\frac{3\alpha_s}{4\delta R^6}2(3a_1^2)\sum_{n>N}I_{n}^2 
\end{eqnarray}

Next we will obtain a constant
 $C$ providing an upper bound on the integrals
 \begin{equation}
 I_n\le\frac{C}{n^{3/2}}
 \end{equation}
 for asymptotically large $n$, which gives us a bound
\begin{align}
    \sum_{n>N}I_n^2\le C^2\sum_{n>N}n^{-3}\le C^2\int_N^\infty x^{-3}dx=\frac{C^2}{2N^2}
\end{align}
and substituting back, 
\begin{align}
    |\mathcal{R}_N(R)|\le\frac{9\alpha_sa_1^2}{4\delta R^6}\frac{C^2}{N^2}
\end{align}
so the error would scale as $O(N^{-2})$ and convergence with $N$ would be guaranteed.

To see that there is such constant $C$
we expand Eq.~(\ref{ec: I_n}) in the small quantity  $\frac{1}{n}$.

First we need the expansion of the factor  $N_n(a_8)$ with explicit value in Eq.~(\ref{ec: N_n}), 
keeping in mind that, due to the colour algebra, the Coulombic radii satisfy  $a_8=2a_1$, and this gives
\begin{align}
    N_n(a_1)=\frac{a^{-5/2}}{\sqrt{2}}n^{-9/2}(1+O(n^{-2}))\ .
\end{align}

Second, we address the sum of Eq.~(\ref{ec: I_n}), performing the expansion with the help of the following equality,
\begin{eqnarray}
    \begin{pmatrix}
        n+3\\
        n-k
    \end{pmatrix}=\begin{pmatrix}
        n+3\\
        k+3
    \end{pmatrix}=\frac{(n+3)!}{(k+3)!(n-k)!}=\nonumber  \\ \frac{(n+3)...(n-(k-1))}{(k+3)!}=\frac{n^{k+3}}{(k+3)!}(1+O(n^{-1}))\ ,  \nonumber  \\
\end{eqnarray}
 by noting that the factor inside that sum 
can be expanded,
\begin{equation}
    (2n/2n+1)^{5+k}=1+O(n^{-1}) 
\end{equation}
and by employing $\Gamma(5+k)=(k+4)!$, 
so that the sum in Eq.~(\ref{ec: I_n})
finally is
\begin{align}
    \sum_{k=0}^{n-2}\frac{(-1)^k}{n^k}\frac{n^{k+3}}{(k+3)!}\frac{(k+4)!}{k!}(1+O(n^{-1}))\\
    n^3\sum_{k=0}^{n-2}(-1)^k\frac{k+4}{k!}(1+O(n^{-1}))\ .
\end{align}
Now, the finite sums may be extended to infinite series up to a controlled error,
\begin{align}
    \sum_{k=0}^{n-2}{\frac{(-1)^k}{k!}}=\sum_{k=0}^{\infty}{\frac{(-1)^k}{k!}}+O(n^{-1})=e^{-1}+O(n^{-1})\\
    \sum_{k=0}^{n-2}{\frac{(-1)^k}{k!}k}=\sum_{k=0}^{\infty}{\frac{(-1)^k}{k!}k}+O(n^{-1})=-e^{-1}+O(n^{-1})
\end{align}
so that the sum becomes
\begin{align}
    3n^3e^{-1}(1+O(n^{-1})) \ .
\end{align}
Taking this result back to Eq~(\ref{ec: I_n}):
\begin{align}
    I_n&=\frac{3\sqrt{2}}{e}a_1(1+O(n^{-1})) 
\end{align}
we see that, for $n>>1$,
\begin{equation}
    I_n\sim Cn^{-3/2}\ , \ \  \ \ \ C={3\sqrt{2}}e^{-1}a_1 \ .
\end{equation}

Thus, the error is finally bound from above as follows
\begin{align}
\boxed{
    |\mathcal{R}_N(R)|\le\frac{81}{2e^2\delta}\frac{a_1^4\alpha_s}{R^6}\frac{1}{N^2}}
\end{align}
and therefore  $R_N(R)=O(N^{-2})$.

Once convergence has been demonstrated, we have implemented computer codes to numerically compute the integrals $I_n$.

By plotting the value of the constant $C$ as a function of 
 $a_1$, the radius of the colour-singlet glueball, 
and performing a linear fit in the asymptotically large regime
$n\in(2000,5000)$, we obtain the constant  $C_{num}=1.563$, which is in great agreement with the theoretical value $C=3\sqrt{2}e^{-1}\approx1.561$, with a very small relative error
\begin{align}
    \frac{|C-C_{num}|}{C}\times 100\approx0.14\%\ .
\end{align}

We thus conclude that truncating the series of the potential for the purpose of a numerical calculation is perfectly controlled.

\section{Estimate including the linear potential}
\label{app:includelinear}
We now proceed to an estimate including the linear potential among colour charges, in addition to the Coulomb one, 
to make sure that the presence of this flux-tube induced phenomenon germane to colour confinement will not qualitatively change our statement about the dominance of the colour Van der Waals force in glueball interactions.
Equation~(\ref{Poteff}) in the main text, the effective Van der Waals potential for the complete underlying Cornell one, 
can be rewritten in terms of the tensors $P_{kl}$ and $Q_{kl}$ defined in Eq.~(\ref{projector}) and~(\ref{skewedprojector}) taking the form
\begin{equation}
    V_{eff}(R)=\sum_{|\chi_1\chi_2\rangle\neq|11\rangle}
          \frac{(aP_{ij}+\alpha_s \frac{Q_{ij}}{R^2})
                (aP_{kl}+\alpha_s \frac{Q_{kl}}{R^2}) d^{ij}d^{kl}
                }{2E_1-(2E_{\chi}+C_2(\chi)(aR-\alpha_sR^{-1}))}
                \label{Veffwithlin}
\end{equation}
with the matrix element of two dipoles along arbitrary Cartesian axes,
\begin{equation}d_{kl}=\langle11|D^{A,k}_{(1)}D^{A,l}_{(2)}|\chi_1\chi_2\rangle\ .
\end{equation}

After summing over glueball polarizations, we obtain
\begin{eqnarray}
    \sum_{m_1,m_2}\langle11|D_{(1)}^{Ak}D_{(2)}^{Al}|\chi_1\chi_2\rangle\langle11|D_{(1)}^{Ai}D_{(2)}^{Aj}|\chi_1\chi_2\rangle\nonumber \\
    =\frac{1}{3}\delta^{ki}\delta^{lj}|\langle11|\boldsymbol{D}_{(1)}^A\cdot\boldsymbol{D}^A_{(2)}|\chi_1\chi_2\rangle|^2\ .
    \label{sumpolswithlin}
\end{eqnarray}
Employing this Eq.~(\ref{sumpolswithlin})
with Eq.~(\ref{Veffwithlin})
and simplifying the tensor contractions with
\begin{eqnarray}
P_{kl}P^{kl} &=& \frac{2}{R^2} \nonumber \\
Q_{kl}Q^{kl}&=&\frac{6}{R^2} \nonumber \\
Q_{kl}P^{kl}&=&\frac{2}{R^2}
\end{eqnarray}
it follows  that the effective potential 
becomes
\begin{eqnarray}
    V_{\rm eff}(R)=\frac{1}{3}\left(\frac{2a^2}{R^2}+\frac{6\alpha_s^2}{R^6}+\frac{4a\alpha_s}{R^4}\right)
    \nonumber \\
    \sum_{|\chi_1\chi_2\rangle\neq|11\rangle}
          \frac{|\langle11|\boldsymbol{D}^A_{(1)}\cdot\boldsymbol{D}^A_{(2)}|\chi_1\chi_2\rangle|^2}{2E_1-(2E_{\chi}+C_2(\chi)(aR-\alpha_sR^{-1}))}\ . \nonumber \\ 
          \label{eq: Poteffgeneral}
\end{eqnarray}
To proceed we would need the glueball wavefunctions in the singlet and in the coloured {\it e.g.} antisymmetric octet representations; but in the presence of a linear potential, technically the energy of the coloured glueballs is infinite (a manifestation of confinement) so that the eigenvalue problem has to be treated very carefully in the context of a larger system. 

\begin{figure}
    \centering
    \includegraphics[width=0.9\linewidth]{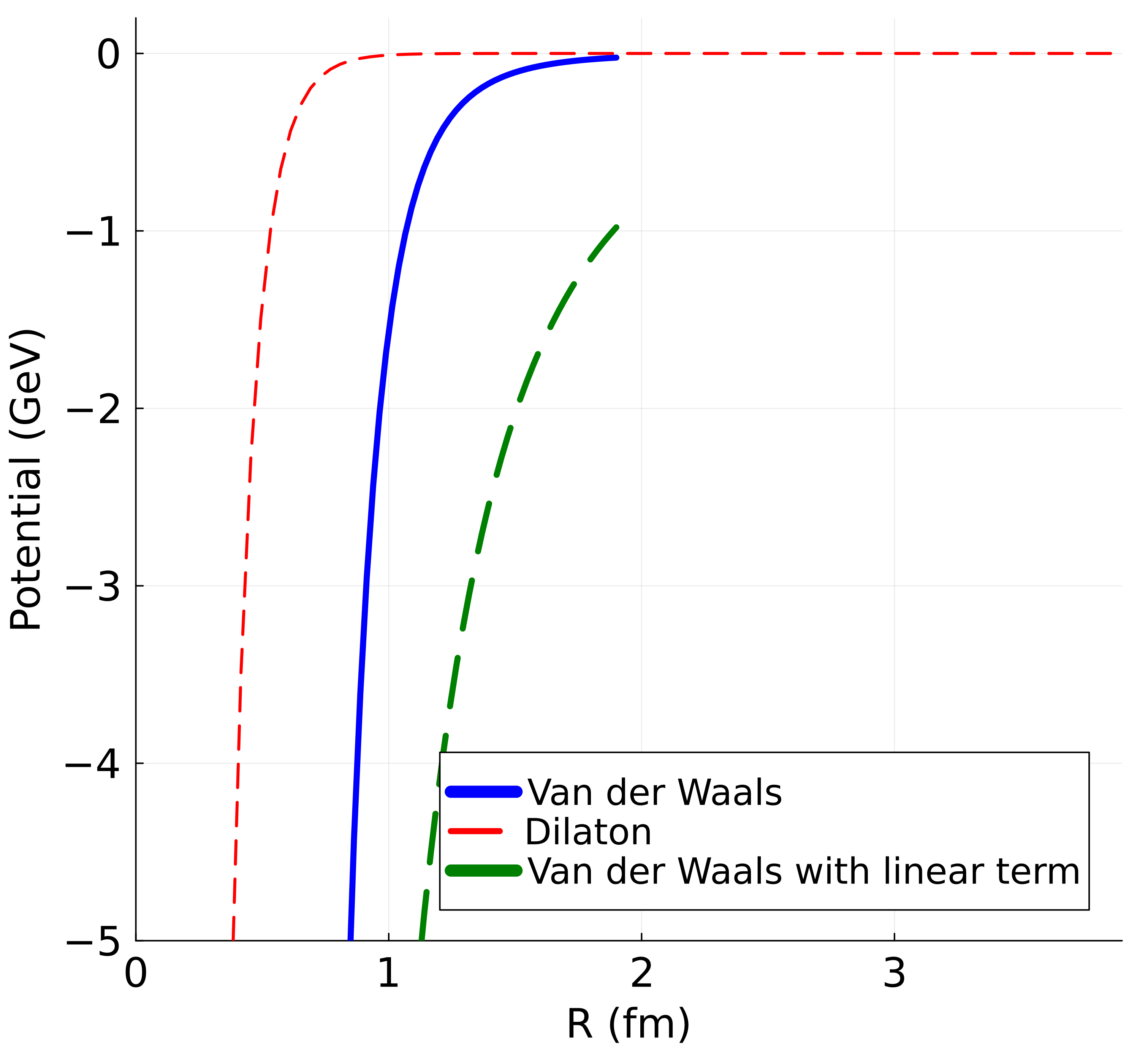}
    \caption{We add to Fig.~\ref{fig:comparison} a third line (dashed, fern green online) in which the contribution of the underlying linear potential has been estimated from Eq. (\ref{eq: Poteffgeneral}) together with the Coulomb one. The resulting effective potential is substantially larger (in modulus) than the Coulomb-only computation, and corroborates our observation that the Van der Waals potential is a very important contribution to the glueball-glueball interaction.
    \label{fig:withlin}}
\end{figure}

For a quick estimate we will proceed employing the Coulombic wavefunctions of Appendix~\ref{app:wfsCoulomb} to get the matrix elements with the linear potential. This is formally a perturbative or also a variational approximation (depending on the point of view), which {\it a posteriori} we know to not be very good (see Fig.~\ref{fig:withlin}) since the contribution of the linear potential is very large, but at least it shows that the main point of the article, that the Van der Waals contribution to glueball-glueball interactions is large, happens to be sound.

Under this variational approximation Eq. (\ref{eq: Poteffgeneral}) takes the form
\begin{eqnarray}
    V_{eff}(R)=\frac{3}{8}\left(\frac{2a^2}{R^2}+\frac{6\alpha_s^2}{R^6}+\frac{4a\alpha_s}{R^4}\right) \sum_{n_1,n_2\ge2}^\infty
    \nonumber \\ 
          \frac{I_{n_1}^2I_{n_2}^2}{-9\mu\alpha_s^2(1-(2n_1)^{-2}-(2n_2)^{-2})-3(aR-\alpha_s/R)}\ ,\nonumber \\
          \label{eq: Poteffgeneral}
\end{eqnarray}
for whose evaluation all ingredients are already at hand. This formula is the one  directly implemented to yield Fig.~\ref{fig:withlin}.

\newpage 
\section{Extension to $SU(2)$}
For ease of comparison with lattice computations of the glueball-glueball scattering, where the extant data seems to have been computed within $SU(2)$, we are going to extend in this appendix 
the calculation of the Van der Waals potential here presented 
to the smaller $SU(2)$ gauge group. This is of little relevance for Chromodynamics, but may be of some interest to constrain specific dark matter sectors. 
For this smaller group quarks have two colours only, and therefore gluons $|a\rangle$ populate a three-dimensional representation of the group, with $a=1,2,3$, identical to the spin group.

Each glueball may be found in a spin 0, 1 or 2 representation,
\begin{equation}
    3\otimes 3=1\oplus 3\oplus 5\ .
\end{equation}
The overall two-glueball system must however be in a singlet contained in the decomposition of the product $(3\otimes 3)\otimes (3\otimes 3)$ into irreducible representations. 
The only ones that include a singlet are $1\otimes 1$, $3\otimes 3$ and $5\otimes 5$. We recall that the $1$ and $5$ states of a glueball are symmetric under exchange of its two gluons, whereas the $3$ state is antisymmetric. Using the index notation $|j_i,m_i\rangle$ $i=1,2$ of $SU(2)$ for each glueball, the colour singlet states of each representation are:
\begin{align}
    |c_3,c_3\rangle&=\frac{1}{\sqrt{3}}\left(|1,1\rangle|1,-1\rangle-|1,0\rangle|1,0\rangle+|1,-1\rangle|1,1\rangle\right)\\
    |c_5,c_5\rangle&=\frac{1}{\sqrt{5}}\left(|2,2\rangle|2,-2\rangle
    -|2,1\rangle|2,-1\rangle\right. \nonumber \\ 
    &\left. +|2,0\rangle|2,0\rangle-|2,-1\rangle|2,1\rangle+|2,-2\rangle|2,2\rangle\right).
\end{align}
Or in cartesian notation (which we will adopt):
\begin{align}
    |c_3,c_3\rangle&=\frac{1}{\sqrt{3}}\delta^{ab}|c_3;a\rangle|c_3;b\rangle\\
    |c_3,c_3\rangle&=\frac{1}{\sqrt{5}}\delta^{ab}|c_5;a\rangle|c_5;b\rangle,
\end{align}
with $a=1,2,3$.
Now, as in the case of $SU(3)$, since the 5-plet states are symmetric under exchange of two gluons, they do not contribute to the matrix element  of the colour dipole operator from the singlet-singlet glueball state, so the only possible polarization is from singlet to triplet.\\

Let us again consider bound states of the Coulomb potential with a colour factor, in total analogy with $SU(3)$, but now in the case of $SU(2)$ the potential becomes
\begin{equation}
    V(r,\chi)=-\frac{\alpha_s}{r}\left(2-\frac{C_2(\chi)}{2}\right)
\end{equation}
where the factor $2$ in the first term is the Casimir of the adjoint representation of $SU(2)$. Thus, the potential of the gluons in a glueball in the different representations of $SU(2)$ is:
\begin{align}
    V(r,1)&=-\frac{2\alpha_s}{r}\\
    V(r,3)&=-\frac{\alpha_s}{r}\\
    V(r,5)&=\frac{\alpha_s}{r}.
\end{align}
We note that for the representation $5$ the potential is repulsive, in accordance with the fact that this representation does not contribute. Therefore, we can assign an effective coupling constant $\alpha_{eff}(\chi)=\alpha_s(2-C_2(\chi)/2)$ to each representation, and the Bohr radius for a bound state is again given by the equation:
\begin{equation}
    a_\chi=\frac{1}{\mu\alpha_{eff}(\chi)}=\frac{2}{m_g\alpha_{eff}(\chi)}.
\end{equation}
With this expression, the Bohr radii are:
\begin{align}
    a_1&=\frac{1}{m_g\alpha_s}\\
    a_3&=\frac{2}{m_g\alpha_s}
\end{align}
so that the relative relation $a_3=2a_1$ is again satisfied and therefore the radial integrals will remain the same, except for changing the value of the parameter $a_1(SU(3))$ to $a_1(SU(2))$.\\

Let us obtain the normalizations of the colour states of a glueball as in $SU(3)$:
\begin{align}
    |c_1\rangle&=N_1\delta^{ab}|a\rangle|b\rangle\\
    |c_3;\lambda\rangle&=N_3\epsilon^{ab\lambda}|a\rangle|b\rangle,
\end{align}
imposing $1=\langle c_1|c_1\rangle$ and $\delta^{\lambda\mu}=\langle c_3;\lambda|c_3;\mu\rangle$ one obtains:
\begin{align}
    N_1&=1/\sqrt{3}\\
    N_3&=1/\sqrt{2}.
\end{align}
As in $SU(3)$ we factorize the colour part of the matrix element in terms of the colour generators of each gluon $T^A_aT^A_b$ with $a=1,2$ $b=3,4$. We calculate the following, and as in $SU(3)$ with it we have all of them:
\begin{align}
    \langle c_1c_1|T^A_1T^A_3|c_3;\lambda,c_3;\mu\rangle&=\langle c_1|T^A_1|c_3;\lambda\rangle\langle c_1|T^A_3|c_3;\mu\rangle.
\end{align}
For this it is enough to calculate one factor,
\begin{align}
     \langle c_1|T^A_1|c_3;\lambda\rangle&=\frac{1}{\sqrt{6}}\delta^{ab}\epsilon ^{cd\lambda}\langle a |T_1^A|c\rangle\langle b|d\rangle\\
     &=\frac{(-i)}{\sqrt{6}}\delta^{ab}\epsilon ^{cd\lambda}\epsilon^{Aec}\langle a |e\rangle\langle b|d\rangle\\
     &=\frac{(-i)}{\sqrt{6}}\delta^{ab}\epsilon ^{cd\lambda}\epsilon^{Aec}\delta_{ae}\delta_{bd}\\
     &=\frac{(-i)}{\sqrt{6}}\epsilon^{ca\lambda}\epsilon^{Aac}=\frac{2i}{\sqrt{6}}\delta^{A\lambda}.
\end{align}
Thus, considering the global singlet state formed by triplet states of glueballs, we have:
\begin{align}
    \langle c_1c_1|T^A_1T^A_3|c_3c_3\rangle&=\frac{1}{\sqrt{3}}\delta^{\lambda\mu}\langle c_1|T^A_1|c_3;\lambda\rangle\langle c_1|T^A_3|c_3;\mu\rangle\\
    &=\frac{1}{\sqrt{3}}\delta^{\lambda\mu}\left(\frac{2i}{\sqrt{6}}\delta^{A\lambda}\right)\left(\frac{2i}{\sqrt{6}}\delta^{A\mu}\right)\\
    &=-\frac{2}{3\sqrt{3}}\delta^{\lambda\mu}\delta_{\lambda\mu}=-\frac{2}{\sqrt{3}}.
\end{align}
And with this the rest of the colour matrix elements are determined, just as in the case of $SU(3)$. The spatial part of the matrix element is identical, except that the integrals $I_n$ (which have the same functional form) have a different value of the parameter $a_1$.\\
Altogether, the matrix element is:
\begin{align}
    \boxed{\langle 11|\boldsymbol{D}_{(1)}^A\cdot\boldsymbol{D}^A_{(2)}|3_A,n_1;3_A,n_2\rangle=\frac{2}{\sqrt{3}}I_{n_1}I_{n_2}} \ .
\end{align}
Taking into account that the energies in $SU(2)$ are:
\begin{align}
    E_1=E_{1,1}&=-2\mu\alpha_s^2\\
    E_{3n}&=-\frac{\mu\alpha_s^2}{2n^2},
\end{align}
and substituting into equation (58) as in $SU(3)$, we obtain an effective potential for the London force:
\begin{eqnarray}
    V_{\rm eff}^{SU(2)}(R)=-\frac{4}{3}\frac{\alpha_s}{R^6}\times\nonumber\\\sum_{n_1,n_2\ge2}^\infty\frac{I_{n_1}^2I_{n_2}^2}{\mu\alpha_s(2-(2n_1)^{-2}-(2n_2)^{-2})-1/R} \nonumber\\
    \label{eq:VdWSU(2)Coulomb}
\end{eqnarray}
which is rather similar to Eq.~(\ref{ec:pot ef coulomb}), except for small numerical factors of order 1.

\clearpage
{}
\end{document}